\documentclass{article}

\usepackage{PRIMEarxiv}

\usepackage[utf8]{inputenc} % allow utf-8 input
\usepackage[T1]{fontenc}    % use 8-bit T1 fonts
\usepackage{hyperref}       % hyperlinks
\usepackage{url}            % simple URL typesetting
\usepackage{booktabs}       % professional-quality tables
\usepackage{amsfonts}       % blackboard math symbols
\usepackage{nicefrac}       % compact symbols for 1/2, etc.
\usepackage{microtype}      % microtypography
\usepackage{lipsum}
\usepackage{fancyhdr}       % header
\usepackage{graphicx}       % graphics
\graphicspath{{media/}}     % organize your images and other figures under media/ folder
\usepackage{amsmath}
\usepackage{stfloats}
\title{Conformal Predictions Enhanced Expert-guided Meshing with Graph Neural Networks}

\author{
  Amin Heyrani Nobari \\
  Department of Mechanical Engineering,\\
  Massachusetts Institute of Technology,\\
  Cambridge, MA, USA\\
  \texttt{ahnobari@mit.edu} \\
  \And
  Justin Rey \\
  Lincoln Laboratory,\\
  Massachusetts Institute of Technology,\\
  Lexington,MA, USA
  \AND
  Suhas Kodali \\
  Lincoln Laboratory,\\
  Massachusetts Institute of Technology,\\
  Lexington,MA, USA
  \And
  Matthew Jones \\
  Lincoln Laboratory,\\
  Massachusetts Institute of Technology,\\
  Lexington,MA, USA
  \And
  Faez Ahmed, \\
  Department of Mechanical Engineering,\\
  Massachusetts Institute of Technology,\\
  Cambridge, MA, USA\\
  \texttt{faez@mit.edu} \\
}

\begin{document}
\maketitle

\begin{abstract}
Computational Fluid Dynamics (CFD) is widely used in different engineering fields, but accurate simulations are dependent upon proper meshing of the simulation domain. While highly refined meshes may ensure precision, they come with high computational costs. Similarly, adaptive remeshing techniques require multiple simulations and come at a great computational cost. This means that the meshing process is reliant upon expert knowledge and years of experience. Automating mesh generation can save significant time and effort and lead to a faster and more efficient design process. This paper presents a machine learning-based scheme that utilizes Graph Neural Networks (GNN) and expert guidance to automatically generate CFD meshes for aircraft models. In this work, we introduce a new 3D segmentation algorithm that outperforms two state-of-the-art models, PointNet++ and PointMLP, for surface classification. We also present a novel approach to project predictions from 3D mesh segmentation models to CAD surfaces using the conformal predictions method, which provides marginal statistical guarantees and robust uncertainty quantification and handling. We demonstrate that the addition of conformal predictions effectively enables the model to avoid under-refinement, hence failure, in CFD meshing even for weak and less accurate models. Finally, we demonstrate the efficacy of our approach through a real-world case study that demonstrates that our automatically generated mesh is comparable in quality to expert-generated meshes and enables the solver to converge and produce accurate results. Furthermore, we compare our approach to the alternative of adaptive remeshing in the same case study and find that our method is 5 times faster in the overall process of simulation. The code and data for this project are made publicly available at https://github.com/ahnobari/AutoSurf Upon paper acceptance.
\end{abstract}

\date{Version \versionno, \today}%% You can modify this information as desired. 
							%% Putting \date{} will suppress any date.  
							%% If this command is omitted, date defaults to \today
							%% This command must come somewhere before \maketitle

\maketitle %% This command creates the author/title/abstract block. Essential!

%%%%%%%%%%%%%%%%%%%%%%%%%%%%%%%%%%%%%%%%%%%%%%%%%%%%%%%%%%%%%%%%%%%%%%%%%%%%%%%%%%%%%%%%%%%%%%%%%%%%%%%
%%%%%%%%%%%%%%%%%%%%%  End of fields to be completed. Now write! %%%%%%%%%%%%%%%%%%%%%%%%%%%%%%%%%%%%%%

\section{Introduction}
\begin{figure*}
\centering\includegraphics[width=\linewidth]{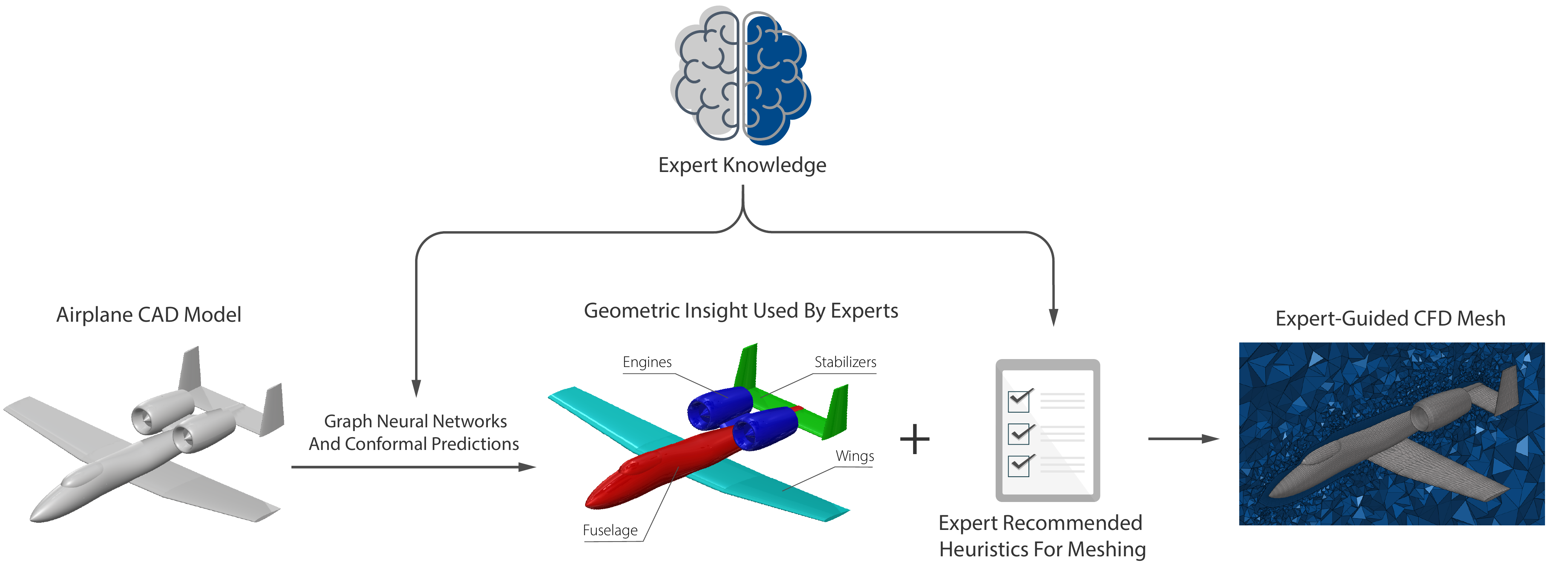}
\caption{Expert-guided Mesh Generation Using Graph Neural Networks and Conformal Predictions}
\label{fig:motivation}
\end{figure*}

Computational Fluid Dynamics (CFD) has revolutionized how engineers analyze fluid flow phenomena, leading to remarkable advances in engineering applications, including aerospace, automotive, and environmental engineering. 
Accurate and reliable simulations of fluid flow phenomena can lead to safer and more efficient designs, reducing the costs of development and improving overall performance. This can positively impact the broader society by enhancing energy consumption efficiency, reducing environmental pollution, and improving transportation safety. Fast CFD simulations may also lead to the advancement of scientific knowledge in fluid dynamics, leading to further breakthroughs in the field. Faster CFD also enables researchers to perform high-fidelity CFD simulations rather than relying on low-fidelity and less accurate simulations, which is often the case in many computational design approaches for applications such as aerodynamic design~\cite{padgan,mopadgan,heyrani2021pcdgan}, turbine design~\cite{langroudi2020modeling}, and much more.
However, accurate simulations of fluid flow require a proper representation of the geometry and the physics of the problem. 

The geometry is usually represented as a discrete set of elements called a mesh, which defines the computational domain and discretizes the equations governing the fluid flow. Therefore, the quality of the mesh has a significant impact on the accuracy and reliability of the CFD simulations. Meshing is the process of creating a finite element mesh. The mesh quality directly affects the accuracy and efficiency of the numerical simulations. 
While highly refined meshes may seem like a logical approach to ensure precision, this method comes with enormous computational costs. Ideally, the mesh should have enough resolution to capture the details of the fluid flow but, at the same time, should be sufficiently coarse to minimize the computational cost. 
Therefore, optimal mesh refinements need to be made, typically in a few critical regions of a structure (e.g., an aircraft's nosecone) that impact the fluid flow around it. However, identifying where to apply refinements to generate a good mesh is an intricate task that depends on the specific geometry and flow conditions. This means that significant time and effort are given to the meshing process as it is crucial to the success of a simulation. In case of a poor mesh quality, the entire time-consuming simulation process needs to be repeated. Consequently, generating good computational meshes often requires years of experience from CFD experts. Thus automating the meshing process is of significant value to the overall CFD process and can save significant time and effort and lead to a faster and more efficient design process.

Meshing has been an active area of research for several decades, and a vast number of mesh generation techniques have been proposed.
An automated meshing algorithm must generate a mesh that satisfies both accuracy and computational efficiency requirements. In automating mesh generation for CFD, significant research has been conducted on adaptive re-meshing methods~\cite{adaptive_remeshing_book,fidkowski2011review}, where an initial mesh is iteratively refined concurrently with the solver by tracking errors in the domain and refining the mesh based on heuristic rules or optimization~\cite{adaptive_remeshing_book,fidkowski2011review,becker_rannacher_2001,BABUSKA19871,eriksson_estep_hansbo_johnson_1995,VERFURTH199467,10.1007/978-3-642-18775-9_13}. However, these iterative adaptive meshing approaches are often slow and incur a high computational cost, making these methods often lead to much longer simulation times which makes them inferior to expert generated meshes that may not be as optimal as adaptive ones~(maybe slightly over refined) but will lead to a valid solution in less time.  

Despite the progress made in this field, meshing remains a challenging task, especially when dealing with complex geometries. 
In recent years deep learning models have shown great promise in many different domains of engineering design~\cite{regenwetter2022deep}. Substantial works have been proposed for improving iterative methods such as adaptive re-meshing using deep learning models~\cite{huang2021machine,FIDKOWSKI2021109957,pfaff2021learning}. However, most of these approaches focus on accelerating adaptive re-meshing and ignore the human experts involved in the mesh generation process.
Our discussions with CFD experts at the Lincoln Laboratories revealed that, in reality, experts often use specific heuristics or thumb rules to generate meshes instead of adaptive meshing, which can be time-consuming. The rules are specifically handy when the underlying geometries do not change significantly from one instance to another. 
Many of these heuristic rules require a deeper understanding of the geometry of the problem. In this paper, we attempt to accelerate this expert-guided meshing process.
% In this paper, we propose a machine learning-based scheme for automating expert-guided mesh generation process for airplane models~(Fig.\ref{fig:motivation}). Our approach leverages Graph Neural Networks~(GNN) to gain a deeper understanding of geometry. This allows us to develop heuristic-based algorithms to automatically generate meshes that match expert-level mesh quality and automate the meshing process. In our approach, we take rough~(computationally inexpensive) surface meshes of airplane models generated by CAD software and develop a GNN-based model to perform mesh segmentation on airplane parts. 
% It is no secret that robust reliability and risk assessment play a major part in aerospace engineering and design. As such, it is important to enable risk assessment and proper uncertainty quantification for models developed in this field as a necessary component.
% With this in mind, we introduce another component to our approach to uncertainty management. By applying the conformal prediction method~\cite{doi:10.1080/01621459.2017.1395341,10.5555/3495724.3496026,angelopoulos2020uncertainty} we are able to not only quantify the uncertainties associated with our model but to provide a mechanism of risk handling with marginal statistical guarantees. This enables designers to set the risk level they find acceptable and ensure that the generated mesh will be at worst over-refined~(with additional simulation cost) in most cases with under-refinement only occurring at a level less than or equal to the acceptable risk.

We present a machine learning-based scheme for automating the expert-guided mesh generation process for airplane models, with the aim of achieving expert-level mesh quality~(Fig.\ref{fig:motivation}). Our approach utilizes Graph Neural Networks (GNN) to gain a deeper understanding of geometry and develop heuristic-based algorithms that automatically generate high-quality meshes. We begin with rough surface meshes of airplane models generated by CAD software and use a GNN-based model for mesh segmentation of airplane parts. In the aerospace industry, reliable risk assessment and uncertainty quantification are essential components of design. To address this, we introduce a component to our approach that utilizes conformal prediction methods~\cite{doi:10.1080/01621459.2017.1395341,10.5555/3495724.3496026,angelopoulos2020uncertainty} to quantify uncertainties associated with our model and provide a mechanism for risk handling with marginal statistical guarantees. By enabling designers to set an acceptable risk level, our approach ensures that the generated mesh is, at worst over-refined, with under-refinement occurring only at an acceptable level of risk. The last step of our approach involves projecting the model predictions onto the CAD surfaces to identify the aircraft's individual parts. Expert-guided heuristic rules are then applied to determine the necessary mesh fidelity and refinements for each part, which are used to automatically generate CFD meshes for each aircraft model~(figure ~\ref{fig:motivation}). We also release a new annotated dataset of aircraft models, which is created by a new  augmenting approach proposed in this paper.

Our research paper provides several significant contributions to the field, including:

\begin{enumerate}

\item \textbf{Application Impact:} We establish an approach for an expert-guided CFD meshing process for aircraft by combining graph-neural networks based predictions with a rule-based approach.

\item \textbf{Method:} We introduce a new 3D segmentation algorithm that outperforms two state-of-the-art models, PointNet++ and PointMLP, for surface classification.

\item \textbf{Robust Risk Management:} We present a novel approach to project predictions from 3D mesh segmentation models to CAD surfaces using the conformal predictions method, which provides marginal statistical guarantees and robust uncertainty quantification and handling.

\item \textbf{Dataset:} We provide a dataset of realistic airplane models with segmentation labels, which can be used to benchmark future 3D segmentation models.

\item \textbf{CFD Case Study:} Through a real-world case study, we demonstrate that our automatically generated mesh is comparable in quality to expert-generated meshes, and enables the solver to converge and produce accurate results, while being more than 5 times faster than the alternative method of adaptive meshing.\end{enumerate}

\section{Background And Related Works}
Our approach is based on 3D model segmentation where we take meshes of aircraft models and predict which part of the aircraft each face in the mesh belongs to. We then map these predictions to CAD surfaces, which are then used for determining mesh settings. In this section, we introduce a brief background on 3D segmentation models and conformal predictions, both of which play a major role in our work.

\subsection{3D Segmentation and Mesh Segmentation}
% With the increased popularity of deep learning-based methods, various learning methods have been proposed in the computer vision and computer graphics communities for 3D shape classification and segmentation. Although conventional algorithms also exist for this purpose, given the significant performance gap between conventional and learning-based methods here, we will only focus on learning-based 3D shape segmentation methods which significantly outperform conventional methods.

% In general, we observe that there are three primary approaches for 3D shape segmentation: 1) Voxelizing 3D shapes and using 3D convolutional neural networks~(CNN) to generate 3D semantic segmentations labels, 2) Using Point cloud representations and applying point-based approaches to classify each point in the point cloud, 3) Using a surface mesh representation of 3D shapes and classifying either vertices, edges or faces of the mesh. In our work we use mesh representation and classify the faces of the mesh, this is the most physically meaningful segmentation as faces represent surfaces of the 3D model which would belong to different parts of an object. In the following sections, we will briefly discuss the state of the art in each of the three approaches to shape classification and highlight some of the relevant parts of these models that we get inspiration from in our method.

The popularity of deep learning-based methods has led to the proposal of various learning techniques in the computer vision and computer graphics communities for 3D shape classification and segmentation. Although conventional algorithms exist for this purpose, we will only focus on deep learning-based methods, as they significantly outperform conventional ones.

There are three primary approaches for 3D shape segmentation: 1) Voxelizing 3D shapes and using 3D convolutional neural networks~(CNN) to generate 3D semantic segmentation labels, 2) Using point cloud representations and applying point-based approaches to classify each point in the point cloud, 3) Using a surface mesh representation of 3D shapes and classifying either vertices, edges, or faces of the mesh. Our method uses a mesh representation and classifies the faces of the mesh. Mesh is the most physically meaningful segmentation, with faces representing surfaces of the 3D model that would belong to different parts of an object.

In the following sections, we will briefly discuss the state of the art in each of the three approaches to shape classification and highlight some relevant parts of these models that inspired our method.

\subsubsection{CNN Based Methods}
One straightforward approach for using CNNs in mesh segmentation is to leverage rendered images of 3D objects from different viewpoints and apply existing CNNs that are highly effective in image processing tasks to make predictions on such images. Su et al.\cite{Su2015MultiviewCN} were pioneers in this approach, using a multi-view CNN for shape classification, but their method could not be easily adapted for semantic segmentation. Later research developed a more comprehensive multi-view framework~\cite{kalogerakis20173d} for shape segmentation, where segmentation maps were predicted for each view using CNNs, and the resulting image-level predictions were projected onto 3D models by enforcing label consistency using Conditional Random Fields~(CRF)~\cite{kalogerakis20173d}.

By converting a three-dimensional shape into a binary voxel format, it is possible to create a grid representation similar to that of a two-dimensional image. This approach enables the same CNN operations used for images to be easily extended to three-dimensional grids, allowing for a seamless transfer of traditional image-based techniques to the realm of shapes. Wu et al.~\cite{7298801} were the first to explore this idea for 3D shapes, and subsequent works have expanded on this approach~\cite{Brock2016GenerativeAD,Tchapmi2017SEGCloudSS,Hanocka2018ALIGNetPA,7353481,10.1109/TIP.2020.3039378}. However, volumetric representations and image-based methods demand significant computational resources and extensive memory usage. Furthermore, binary voxel representations do not readily map to more accurate representations of 3D shapes, such as meshes. Given these limitations, our work departs from image-based and volumetric representations as well as CNN-based methods.
\subsubsection{Point cloud Approaches}

Point clouds are a common representation method of 3D shapes that can be easily obtained from various other types of 3D shape representations, leading to significant research efforts in utilizing point-based deep learning methods for 3D shape analysis. One of the seminal works in this area is PointNet~\cite{Qi_2017_CVPR}, which utilizes $1\times 1$ convolutions on point cloud coordinates for 3D shape prediction, as well as introducing the Transformation Network (T-Net) to enable model invariance to geometric transformations such as rotations and translations~\cite{Qi_2017_CVPR}. This concept has proven to enhance learning generalization and can be applied to most representations using point/vertex positions, such as point clouds or meshes. Our approach also employs a variant of T-Net to take advantage of these benefits.

PointNet++~\cite{NIPS2017_d8bf84be}, a follow-up to PointNet, partitions points into groups to better capture local structures, resulting in significant performance improvements~\cite{NIPS2017_d8bf84be}. Other approaches incorporate information from local neighborhoods to perform dynamic updates by computing the similarity between points based on their euclidean distance in the feature space~\cite{dgcnn}. Some models, such as RSNet~\cite{huang2018recurrent}, sort points and treat them as sequences, using Recurrent Neural Networks (RNN) to capture local features. PointCNN~\cite{NEURIPS2018_f5f8590c} expands the convolution concept beyond local grids to a $\chi$-convolution that operates on points located within their respective Euclidean neighborhoods. However, PointCNN is not invariant to point permutation, which is addressed in PointConv~\cite{Wu_2019_CVPR}, extending the notion of convolution to an operator invariant to order.

One of the current best approach, PointMLP~\cite{ma2022rethinking}, proposed at the International Conference on Learning Representations (2022), removes convolution and instead employs simple multi-layer perceptrons (MLP) as operators. It extends the ideas of PointNet and PointNet++ and outperforms all other models in segmentation and classification tasks, including those using mesh representations~\cite{ma2022rethinking}. In the results section, we show how our proposed approach outperforms PointMLP.

Most of these approaches capture local features by hierarchically grouping points and learning group-wise features. However, if the mesh structure is available, its edges~(i.e., adjacency information) can enable local feature extraction without the need for grouping and complex operations. This is the topic of the next section and the primary reason for our use of mesh representations.

\subsubsection{Mesh Based Approaches}

The non-uniform polygonal mesh serves as the foundation for 3D data representation in computer graphics. This approach utilizes a smaller number of large polygons to cover expansive, flat regions while employing a larger number of polygons to capture intricate details. A mesh accurately captures complex structures and distinguishes them from nearby surfaces by explicitly representing the surface topology. As such, meshes are considered one of the most accurate representations of 3D data. Models that utilize meshes directly are able to capture local features as needed, eliminating the issue of locality present in point clouds. Additionally, mesh representations do not suffer from permutation issues since they capture adjacency rather than a set of unordered points, as seen in point clouds. These advantages make meshes the most efficient and promising representation of 3D data for learning.

Numerous approaches have been explored for mesh-based deep learning models, including the popular MeshCNN approach introduced by Hanocka et al.~\cite{Hanocka2019MeshCNNAN}. This model utilizes mesh edges as the basis for its approach and computes features for each edge in the mesh, which are invariant to transformations. Convolution is then applied to these features in the neighborhood of each edge, determined by the edges sharing faces in the mesh. Despite its success, MeshCNN has limitations due to its reliance on the convolution operator, making it necessary to refine or coarsen meshes to a specific number of edges, thereby limiting its broad applicability.

To overcome this limitation, a more generalizable and versatile approach has been recently explored, namely the graph-based approach. In this class of models, graph neural networks (GNNs) are utilized to analyze 3D meshes, which lend themselves to the graph representation, with vertices serving as nodes. As such, GNNs are well-suited to handle meshes, and recent works have focused on this approach~\cite{8984309,9720214,WEN2021106250,Qi_2017_ICCV}. We combine the GNN-based approaches with point cloud-based approaches in our work to present a new architecture that encompasses the benefits of both classes of models. Further details of our methodology are discussed in the following sections. Next, we shift our focus to a new method, which allows us to make predictions with provable guarantees on the error rates.

\subsection{Conformal Predictions} 

%% Add a definition of conformal predictions and why we are talking about it.
Conformal prediction is a framework in machine learning that provides a measure of confidence in the predictions made by a model without any assumptions on the underlying predictive model. It is based on the idea of constructing prediction regions around each point in the input space rather than just making a single-point prediction. These prediction regions can be calibrated to provide a desired level of confidence. This is made possible through the principle of exchangeability of the data~(i.e., the order of the data does not affect the predictions).

Let's consider a classification task where we have $n$ datapoints $\left\{\left(X_i, Y_i\right)\right\}_{i=1}^n$ with features $X_i \in \mathbb{R}^f$ and a classification label $Y_i \in$ $\mathcal{Y}=\{1,2, \ldots, C\}$ with $C$ possible discrete classes in the label space. If we train a typical machine learning classifier that outputs a softmax probability distribution for each class, we would have a vague notion of the model's certainty as these values are just guesses that the model has made. The reality is that there is no guaranteed accuracy/uncertainty that these distributions are reflective of the true distribution $P_{XY}$. If the cost of mistakes isn't high, one can report the single most likely class for any input which comes with no statistical guarantees. However, in our work, each CFD simulation can be time-consuming, and knowing a measure of confidence in the machine learning algorithm can reduce mistakes. When we obtain predictions on mesh faces, we project them onto CAD surfaces. Each CAD surface receives information from many mesh faces associated with it, and given the mesh CFD fidelity is determined by these predictions, it is important to have some statistical guarantees on the accuracy of the predictions. Conformal predictions enable this by constructing prediction sets $\hat{\mathcal{C}}_{\alpha} \subseteq \mathcal{Y}$ with some risk level $\alpha$ for each input rather than a single prediction. Most notably, these prediction sets come with a marginal guarantee of obeying:

\begin{equation}
\mathbb{P}\left[Y\in \hat{\mathcal{C}}_{\alpha}\left(X\right)\right] \geq 1-\alpha
\end{equation}

These approaches enable such a guarantee by running the model on a test set of data that the model has not seen before and calibrating the uncertainty outputs of the model to a threshold that the model must meet for the above statistical guarantee~\cite{10.5555/3495724.3496026}. This kind of statistical guarantee gives us a much better insight into the uncertainty of the predicted label distributions of the deep learning models which are arbitrary guesses. For these kinds of classification problems, many researchers have proposed conformal prediction schemes which can be easily applied to machine learning models~\cite{angelopoulos2020uncertainty,10.5555/3495724.3496026,doi:10.1080/01621459.2017.1395341}. In our works we take the approach of Adaptive Prediction Sets~(APS) proposed by Romano et al.~\cite{10.5555/3495724.3496026} to construct predictions sets which we then use for a voting algorithm.

Making incorrect predictions for CFD meshing can significantly delay the meshing process. By providing a measure of confidence, conformal prediction can help CFD experts make informed decisions and take appropriate action to prevent failures or minimize risks. This is particularly crucial in simulations that may require very large computational resources, which may not be readily accessible at all times. In such cases where failure cannot be tolerated, conformal predictions can be applied with very low-risk acceptance, while in cases where failure can be tolerated to some extent, this risk can be relaxed in favor of a less conservative mesh which can speed up the simulation.

\section{Methodology}
\begin{figure*}[h]
\centering\includegraphics[width=\linewidth]{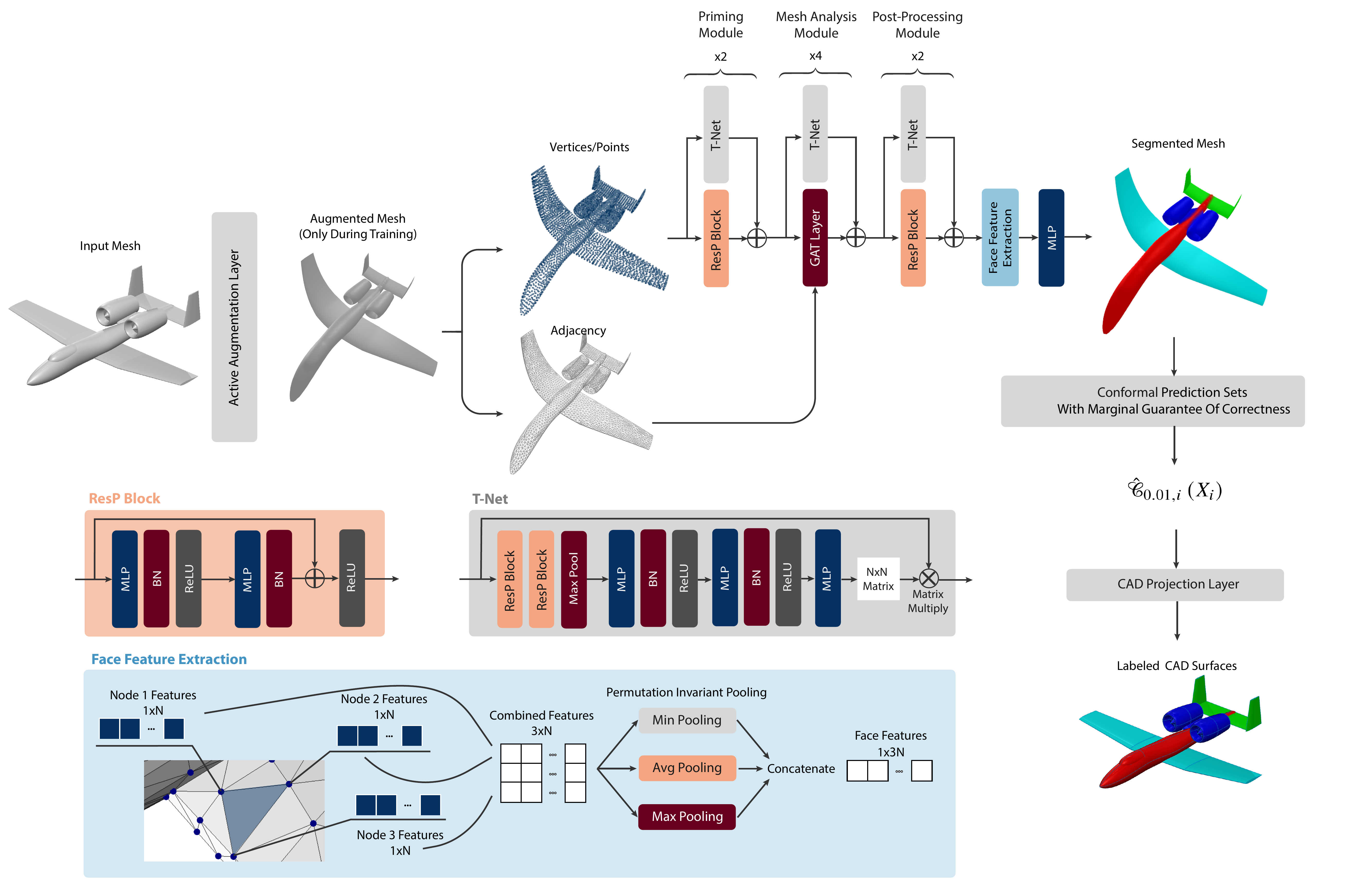}
\caption{The proposed mesh segmentation model is a combination of point-based and graph neural network (GNN)-based approaches. During training, each input first goes through an augmentation layer. Then the augmented mesh is processed through two ResP blocks, followed by four GAT layers and another two ResP blocks, with a T-Net residual applied at each layer. The resulting features for each face are extracted and passed through a classifier MLP to generate segmentation labels for the mesh. Once the mesh is segmented, mesh faces are projected onto CAD surfaces based on conformal prediction sets, which provide marginal statistical guarantees of accuracy.}
\label{fig:architecture}
\end{figure*}

In this paper, we introduce a new model for mesh segmentation which is designed to work with very small datasets without overfitting while having a complex and deep architecture that takes aspects of GNN-based models as well as point cloud-based models to construct a hybrid model with greater accuracy. The overall architecture of our model can be seen in figure ~\ref{fig:architecture}. We then propose an approach to take the predictions on the surface mesh and project them to the CAD surfaces, which can then be used to construct the final CFD mesh. In the sections that follow we will describe the details of our approach.

\subsection{Mesh Segmentation Approach}
In our model, we focus on developing large models that can be trained on very small datasets. This is particularly important in many engineering design applications as usually the size of datasets in engineering design is very small, which makes the application of large deep learning models difficult~\cite{regenwetter2022deep}. However, it has been demonstrated that proper data augmentation can overcome this challenge and enable the application of large deep-learning models without overfitting~\cite{stylegan3,regenwetter2022deep,10.1115/1.4052442}. Given this, we introduce an active augmentation layer to our model to enable better generalization of learning and training with few samples. We describe in detail what this layer does and how it is implemented in training in later sections.

After the augmentation, we introduce a novel architecture~(figure ~\ref{fig:architecture}) for mesh face classification using a combination of point-based and graph-based deep learning approaches which combine the effectiveness of point-based models and the versatility of mesh graph-based models to allow for easy and accurate mesh segmentation. 

Finally, we use the conformal predictions approach known as Adaptive Prediction Sets~(APS) proposed by Romano et al.~\cite{10.5555/3495724.3496026} to construct prediction sets from our segmentation model's output such that a marginal statistical guarantee can be given to the users when it comes to the possible labels each face in the mesh may be associated with.
In the following sections, we describe each of the three aspects of our approach in detail.

\subsubsection{Scheduled Augmentation Layer}

\begin{figure}[h]
\centering\includegraphics[width=0.8\linewidth]{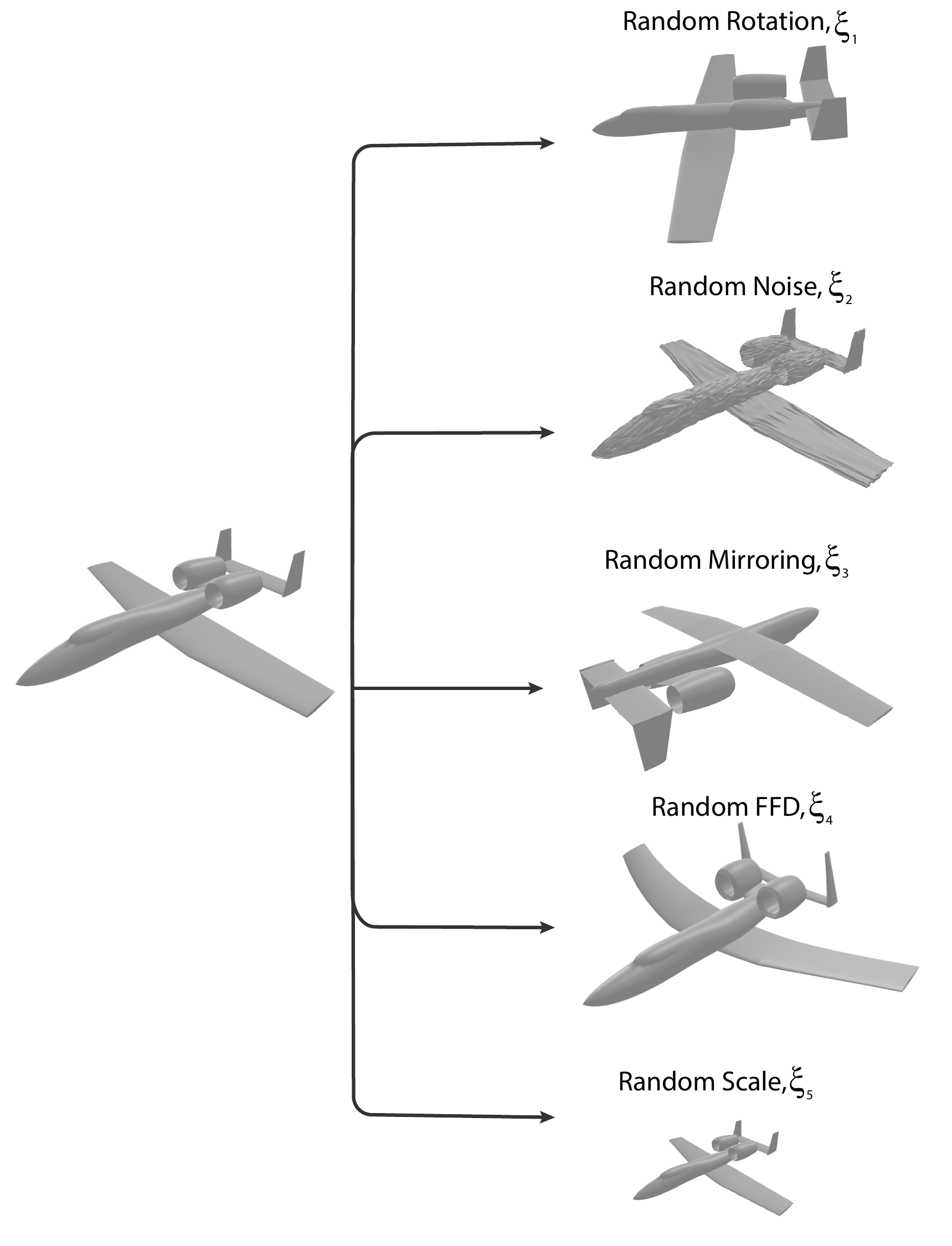}
\caption{Visual demonstration of the five active augmentation approaches we employ in our augmentation layer.}
\label{fig:augmentation}
\end{figure}

When working with small datasets, large deep learning models tend to overfit the training data and fail to generalize to data the model has not been trained on. This leads to a lack of generalization which makes the models unreliable in real-world applications. There are several ways to overcome this problem, one way is to simply gather more data, but that may be too expensive or even impractical. Another way to overcome this issue is to take the input data and alter the data slightly so as to prevent the model to overfit or memorize the data. In computer vision and image data, this is typically done by applying random transformation and noise to the data~\cite{stylegan3}. Here we propose an augmentation layer that applies random changes to the input data to prevent overfitting. 
% Another important thing to note is that if the augmentation is too aggressive it may limit the model's performance by making the task much more difficult and impossible to perform by the model. As such most successful methods either limit the extent of their augmentation~(conventional) or develop adaptive mechanisms for augmentations that adapt to the model's performance and aim to maximize the generalizability as well as possible. One of the first successful implementations of such an approach was the adaptive augmentation mechanism proposed by Karras et al.~\cite{stylegan3} where the authors track the validation performance of their model as training goes on and whenever the model fails to improve they relax the augmentation extent to allow the model to improve on the training data in case this will yield better generalizability in the validation. Similarly, if they observe overfitting in the training~(i.e. training performance increases while validation performance dips) they apply more aggressive augmentation to prevent overfitting. 
In our approach, we introduce a similar method but for 3D data. Our contribution to the augmentation layer is, therefore, two-fold. First, we introduce several augmentation techniques that we find to improve the generalizability of our model. Our second contribution is how we change the extent of our augmentation while the training is going on. In this section, we will first discuss the details of our augmentation pipeline/layer, then we will describe the dynamic augmentation mechanism we use during training.

\paragraph{Augmentation of 3D mesh data:} 
Drawing inspiration from image augmentation methods~\cite{stylegan3}, this paper proposes augmentation mechanisms for 3D mesh data to prevent overfitting in deep learning models. The proposed mechanisms include random rotations in all three directions, adding noise to data by applying random noise to the position of vertices in the mesh, mirroring 3D models with respect to different planes, random application of Free-Form Deformations, and random scaling of the model. One of the common approaches for image augmentation is the random rotation of the images, which is easy to adapt to 3D data by randomly rotating 3D objects in space. Unlike image-based approaches that apply one rotation angle, we apply random rotations in all three directions. Another common augmentation technique is adding noise to data, which we also use in our layer by applying random noise to the position of vertices in the mesh. Another augmentation mechanism we employ is mirroring 3D models with respect to the $XY$, $XZ$, or $YZ$ planes randomly. It also observed that warping images could be an effective form of augmentation in computer vision. We translate that idea to 3D, by using a random application of Free-Form Deformations~(FFD) as analogous to the wrapping technique used for images. Finally, we have random scaling of the model, which either enlarges or shrinks a model slightly. These augmentation mechanisms are all demonstrated in Fig.\ref{fig:augmentation} to help visualize how they work. However, as we discussed earlier, we intend to control these augmentation mechanisms during training. As such, each augmentation mechanism will be controlled by a specific parameter.

For rotation, we randomly sample a 3D angle, $\Theta$, from a uniform distribution of $\Theta \sim U(0,\xi_1) \in \mathbb{R}^3$ and the control parameter $\xi_1$ determines the extent of augmentation. Similarly, for the noise, we sample random noise for each vertex~($v_i$) from a uniform distribution $\epsilon_i \sim U(0,\xi_2) \in \mathbb{R}^3$ and the control parameter $\xi_2$ determines the extent of augmentation. For mirroring, we apply the mirroring across each of the 3 $XY$, $XZ$, and $YZ$ planes randomly with a probability of  $0\leq \xi_3<1$, which controls how likely the mirroring is to occur during augmentation. For warping, we apply the FFD by deforming the model randomly in all directions, and we randomly sample the extent to which these deformations occur from a uniform distribution $U(-\xi_4,\xi_4)$, where $\xi_4$ determines how aggressively the model is deformed. Finally, for scaling, we apply a uniform scale adjustment across all three dimensions, which is sampled from a uniform distribution $SA\sim U(-\xi_5,\xi_5)$, and the model is scaled by $1-SA$. The $\xi_1$, $\xi_2$, $\xi_3$, $\xi_4$, and $\xi_5$ parameters control how aggressively the models are augmented which allows us to control the extent of augmentation during training. More details can be found in the authors' code and data, which will be made available upon acceptance of the paper.

\paragraph{Augmentation Schedule:} When training starts, we start with no augmentation and based on the target value of the $\xi_1$, $\xi_2$, $\xi_3$, $\xi_4$, and $\xi_5$ parameters the augmentation increases as training goes on. In general, we have an augmentation schedule that increases the extent of augmentation as every epoch goes on. The following equation describes this schedule:

\begin{equation}
    \xi_i^T = \xi_i^\infty \times \frac{\left(1+cos\left(\frac{T\pi}{\tau}-\pi\right)\right)}{2}
\end{equation}

where $T$ is the current epoch of training and $\xi_i^\infty$ is the target value of the augmentation parameter while $\xi_i^T$ is the adjusted augmentation parameter at epoch $T$ and $\tau$ is the scaling factor which is set to the maximum number of epochs~(the total number of epochs that we plan to train the model).

% During training, as the augmentation is increased it is expected that, at some point, the model will overcome overfitting. As such, during training, we monitor the validation accuracy at the end of each epoch, and if the validation accuracy is maximized, we checkpoint the model. Simultaneously, the goal is not to make the task too difficult to confuse the model. Consequently, at the same time, as augmentation increases, we reduce the learning rate of the optimizer to allow for a smoother transition to the more challenging task and throw the model too far from the optimal parameters it has established. By doing this, we balance augmentation for generalizability and not hindering the model's learning too much.

While training the deep learning model, we increase the level of augmentation gradually until the model no longer overfits the data. To ensure this, we keep track of the validation accuracy at the end of each epoch and save checkpoints of the model's weights when the accuracy is maximized. However, we do not want to make the task too difficult for the model, so we simultaneously decrease the learning rate of the optimizer as the level of augmentation increases. This helps the model adjust smoothly to the more challenging task and prevents it from straying too far from the optimal parameters it has learned. Thus, we strike a balance between augmentation for generalizability and not hindering the model's learning. More implementation details can be found in our code and data, which will be available to the public upon acceptance of this paper.

\subsubsection{Mesh Segmentation Model}
% As we saw in prior sections point cloud-based models are very popular and successful for 3D shape segmentation and classification. Moreover, these models seem to outperform~\cite{ma2022rethinking} the latest and best mesh-based and purely GNN-based models. However, these models ignore the truest and most versatile representation of 3D data which is the mesh, and rather than extracting the local relationships using the adjacency information captured by the mesh they use neighborhood grouping and sampling to capture local relationships. Despite this, even when models are made to be best suited for specific kinds of tasks the latest point cloud-based model PointMLP~\cite{ma2022rethinking} seems to be able to match their performance closely~\cite{8984309,9720214}. This means we cannot ignore the great performance benefits of point cloud-based methods despite their limitations. As such in our approach, we take the lessons learned from point cloud-based models and combine them with GNN approaches to create a hybrid of the two, which hopefully will encompass the benefits of point cloud models and the versatility of the GNN models.

As discussed in previous sections, point cloud-based models have shown great success in 3D shape segmentation and classification, even outperforming the latest and best mesh-based and GNN-based models~\cite{ma2022rethinking}. However, these models neglect the most accurate and flexible representation of 3D data\cite{Hanocka2019MeshCNNAN}, which is the mesh. Instead, they use neighborhood grouping and sampling to extract local relationships. Despite this, even when specialized for specific tasks, the latest point cloud-based model PointMLP~\cite{ma2022rethinking} is comparable in performance to GNN-based models~\cite{8984309,9720214}. Therefore, we cannot overlook the performance advantages of point cloud-based methods despite their limitations. To create a hybrid model that combines the strengths of both, we incorporate the lessons learned from point cloud-based models into GNN approaches, aiming to achieve the benefits of point cloud models and the versatility of GNN models.

Figure ~\ref{fig:architecture}, shows the overall architecture of our model. There are several stages in our architecture, which we discuss in more detail here.

\paragraph{Priming Module:} As it can be seen, the model starts with a priming module which includes an adaptation of the pointMLP's Residual Point (ResP)~\cite{ma2022rethinking} block with the addition of T-Nets proposed by PointNet~\cite{Qi_2017_CVPR}. As we mentioned in the background section, the T-Net is a mechanism for transformation invariance, which is important when it comes to 3D data, and this matter was taken care of in PointNet++~\cite{Qi_2017_ICCV} and PointMLP~\cite{ma2022rethinking} by grouping and sampling through their geometric affine modules. Since we intend to use the mesh directly in our model and do not wish to perform geometric grouping and sampling, we introduce the T-Net. Our implementation of T-Net is shown in Figure~\ref{fig:architecture}. It takes as input a point set and using a few ResP layers, determines a linear transformation matrix for the point set which is then applied to the point set's features. 

In this module, despite only having point-based components, we do not down-sample the vertices in the mesh to a specific number of points, as this would not allow us to apply GNNs to the entire mesh effectively. 
One of the main limitations of the PointBased models is that during training, only subsets of all the points will be seen, while in our model, we not only input all the points but make classifications on all of our mesh faces. Another limitation of the point cloud-based model is that predictions will be made at the points, not surfaces. Furthermore, predictions in point cloud-based models are made at the points, while true labels are only meaningful for mesh faces in 3D segmentation. Thus, for point-based models, less accurate labels must be determined for individual points, and combining vertices of the mesh is not feasible, as all the vertices associated with faces may not have been captured during the subsampling of training. Therefore, by inputting all the vertices and using GNNs, we avoid this issue and properly analyze the entire model.

\paragraph{Mesh Analysis Module:} Once the data goes through the point-based priming module, we take the features from this stage as the input features for the mesh analysis module. In the mesh analysis module, we represent our data as graphs with node features from the priming module and adjacency matrix same as the mesh adjacency. In this module, we use a GNN approach called Graph Attention Convolution~(GAT)~\cite{velivckovic2017graph}. For the sake of brevity, we will not discuss the details of GAT here and refer the readers to the original body of work for more detail~\cite{velivckovic2017graph}. Simultaneously, we find applying the T-Net from the priming module helps with the model generalization. For the GAT layers, we use 8 attention heads with 64 features each and apply a 10\% dropout during training.

\paragraph{Post-processing Module:} After the mesh analysis module, the resulting features go through the post-processing module, which is made up of ResP blocks and T-Nets. After that, the features of the mesh vertices are aggregated for each face. That is, the features of the vertices associated with each face are aggregated using an order-invariant face feature extraction function~(see figure ~\ref{fig:architecture}). This aggregation function must be order invariant because if the model sees different permutations of the points associated with a face, the outcome should not change. In our approach, we perform aggregation by min-pooling, average-pooling, and max-pooling and concatenate the three in that order. In this way, the features from all three vertices are captured without loss of information while the order invariance of the aggregation function is maintained. Finally, in the post-processing module, we apply an MLP to each face's features to predict the label for each face.

During training, we use the categorical cross-entropy loss to train the model for segmentation:

\begin{equation}
L_{CLS}=-\frac{1}{N} \sum_i^N \sum_j^M y_{i j} \log \left(\hat{y}_{i j}\right)
\end{equation}

where $L_{CLS}$ is the classification loss that we minimize during training and $N$ is the number of faces in a batch during training and $M$ is the number of possible labels, and $y_{i j}$ is the true label for the face $i$ and class $j$ while $\hat{y}_{i j}$ is the label predicted by the model. Beyond this loss, we also apply regularization loss to the transformations predicted by every layer of T-Net we have in our model. This is needed as the dimensionality of the features in the model increase, and the predicted transformations become much larger, which leads to overfitting and instabilities in the model. To mitigate this issue, we apply a regularization loss to the transformations predicted by T-Nets:

\begin{equation}
L_{\text {T-reg}}=\sum_i^N\left\|I-A_i A_i^T\right\|^2
\end{equation}

Where $L_{\text {T-reg}}$ is the regularization loss minimized and $N$ is the number of T-Net layers and $A_i$ is the transformation predicted by the $i$-th T-Net layer. The total loss that we minimize during training is, therefore:

\begin{equation}
L= L_{CLS} + \gamma \times L_{\text {T-reg}}
\label{eqn:5}
\end{equation}

Where gamma is the weight given to the regularization loss.

\subsubsection{Conformal Predictions}
Conformal prediction is a powerful technique for improving the reliability and robustness of machine learning models, making it a useful tool in a wide range of engineering applications.
At this point, we have described a standard 3D segmentation model. However, the performance of deep-learning models is dependent on the data, and they simply make point predictions that are best guesses. The predictions for new test cases do not come with any statistical guarantees. This means that if we only base our classification on the model's best guess, we cannot guarantee any accuracy. In general, uncertainty quantification with any kind of guarantee based on learning is difficult and most work in this area is restricted to Gaussian Processes. However, we could easily overcome this issue in a model-agnostic way by using conformal predictions. In our work, we use the approach proposed by Romano et al.~\cite{10.5555/3495724.3496026} to construct prediction sets that are marginally guaranteed to include the correct prediction 95\% of the time. More formally:

\begin{equation}
\mathbb{P}\left[Y_i\in \hat{\mathcal{C}}_{0.05,i}\left(X_i\right)\right] \geq 0.95
\end{equation}

Where $\hat{\mathcal{C}}_{0.05,i}$ is the prediction set for the input sample~(mesh face) $X_i$ and $Y_i$ is the true label. We then use these prediction sets to map the predictions on the mesh to the CAD surfaces. One thing to note here is that if the acceptable risk level~(i.e., $0.05$) were to be lowered, this may lead to the model becoming more conservative, leading to more surfaces having uncertain labels associated with them, which can lead to an over-refined mesh more often, leading to more expensive CFD simulation. As such, great care must be given to defining an appropriate risk level. In this work, we found that for our model a $0.05$ risk led to no increase in the number of mistakes by the model but correctly enabled the identification of uncertainties and removed any under-refinements.

\subsection{Mapping Mesh Segmentation to CAD Segementation}
\label{sec:cadseg}
\begin{figure*}[h]
\centering\includegraphics[width=0.9\linewidth]{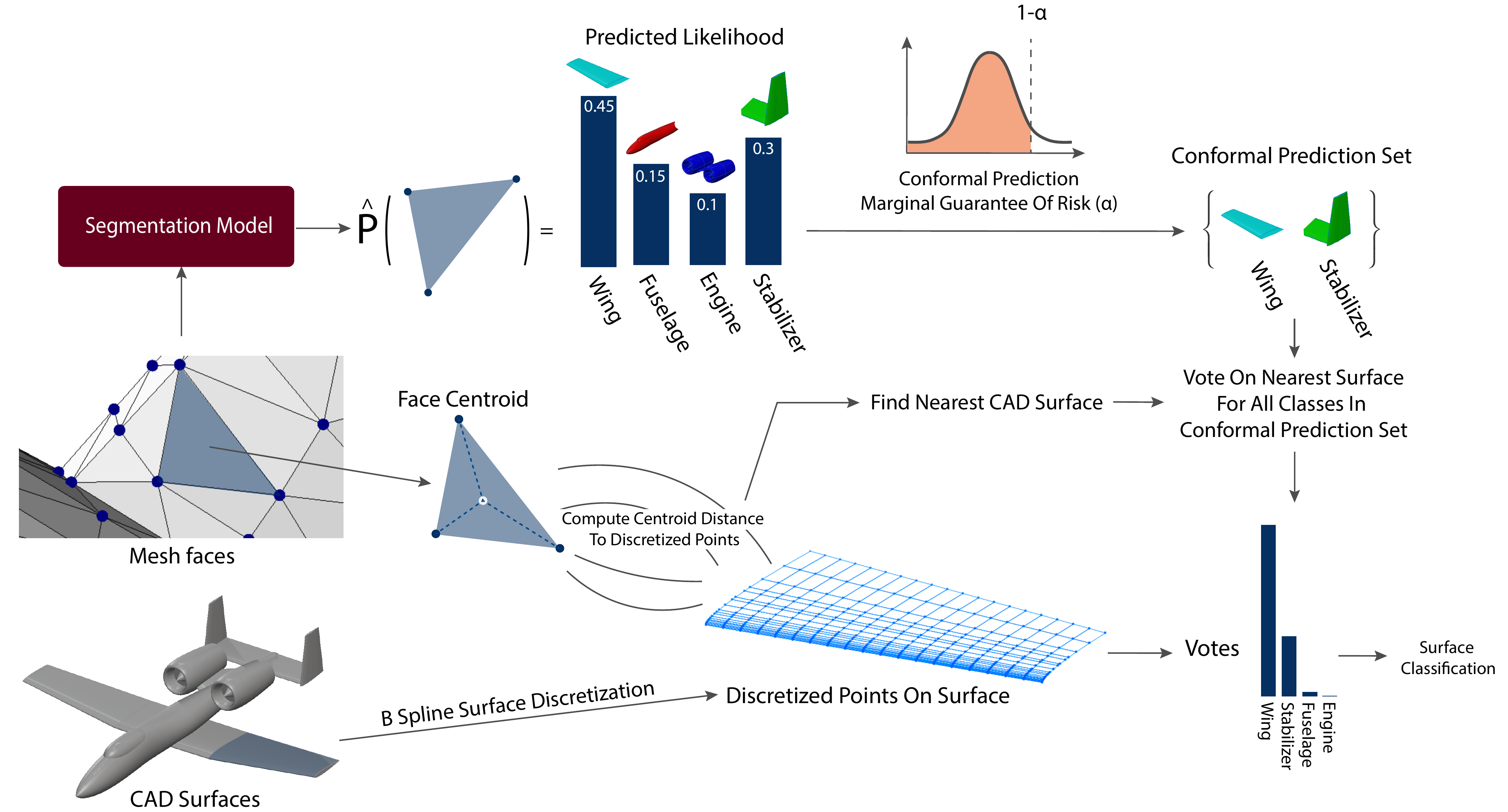}
\caption{Our approach for mapping model predictions to CAD surfaces with marginal statistical guarantees using conformal predictions.}
\label{fig:surface_conformal}
\end{figure*}

% Given that effective CFD meshing requires specific mesh settings to be applied to different surfaces of an airplane model despite classifying different parts of an aircraft in a mesh we need to mesh each surface of the aircraft with specific settings. Therefore, the only practical way to automate the CFD meshing would be to take classifications from the mesh and map them to CAD surfaces which then can be meshed in a CFD meshing platform based on what type of surface it is.

In order to generate an effective Computational Fluid Dynamics (CFD) mesh for an airplane model, it is necessary to apply specific mesh settings to different surfaces of the aircraft. As a result, it is not sufficient to classify different parts of the airplane in a single mesh. Instead, each surface of the aircraft must be meshed with specific settings. Thus, the most practical approach to automating the CFD meshing process would involve mapping the classifications from the mesh to the corresponding surfaces in a CAD model. These surfaces can then be meshed in a CFD meshing platform using the appropriate settings for each surface type.

\paragraph{Estimating Mesh to CAD Surface Distance:} To map predictions from the mesh to the CAD model, we take the predictions for each face in the mesh associated with the CAD files and calculate the location of the centroid of the face. Then we measure the distance from the centroid of each face to all the CAD surfaces and identify the closest surface in the CAD for each mesh face. To measure the distance to different CAD surfaces we take the B-Spline surfaces of the CAD file and discretize the surfaces numerically using the control points defining the B-Spline surface and obtain a grid of points in each surface. Then we measure the distance between any given face centroid and all of the points in the discretized surface and take the minimum of all these distances as the distance from a face centroid to any given CAD surface:

\begin{equation}
    D_{i j} = \min_k \|x_i-x_{j k}\|
\end{equation}

Where $D_{i j}$ is the distance between face $i$ and surface $j$ and $x_i$ is the location of the centroid of face $i$ and $x_{j k}$ is the location of the $k$-th discretized point of the $j$-th surface.

% \paragraph{Voting Mechanism for CAD Surface Mapping:} After the appropriate CAD surface for each face is determined each face will vote on the nearest CAD surface. The voting is done based on the prediction sets from the conformal predictions we described prior with a 95\% marginal guarantee. Then we sum up the votes on each surface, and if only one specific class has achieved majority~(50\%+1 vote) then we pick that prediction as the surface type and if that does not occur we pick the surface classification in the top two highest voted classes that require more refined mesh settings~(see next section ~\ref{sec:meshingrules} details). In this way, at worst we will end up over-refining the mesh on a few surfaces which will still yield a mesh that resolves the flow and leads to valid solutions, however, it comes at a higher computational cost for the solver. Given a very accurate model with a $95\%$ marginal correctness guarantee in the conformal prediction sets, almost always there will be a singular majority, and when this does not occur~(i.e., the conformal prediction cannot guarantee the $95\%$) our model takes the more conservative approach to meshing, avoiding any negative outcome for the most part. This means that if we obtain an accurate enough model we are practically guaranteed to achieve valid mesh settings.

\paragraph{Voting Mechanism for CAD Surface Mapping:} Once the CAD surface for each face is identified, the faces will vote on the nearest CAD surface based on the conformal predictions made previously, which have a 95\% marginal correctness guarantee. The votes for each surface are then tallied, and if a single class obtains the majority vote~(i.e., 50\%+1 vote), it is selected as the surface type. If this does not occur, we select the surface classification from the top two classes with the most votes that require more refined mesh settings~(see section~\ref{sec:meshingrules} for details). This approach ensures that the mesh is not under-refined and can resolve the flow, leading to valid solutions. However, it may result in some surfaces being over-refined, which comes at a higher computational cost for the solver. When the conformal prediction sets have a 95\% marginal correctness guarantee, the model can accurately predict a singular majority in most cases. If this is not possible, our model takes a more conservative approach to mesh to avoid any negative outcomes. Therefore, with an accurate enough model, we can practically guarantee valid mesh settings.

\subsection{Expert-Guided Automated Meshing}
\label{sec:meshingrules}
\begin{figure}[h]
\centering\includegraphics[width=\linewidth]{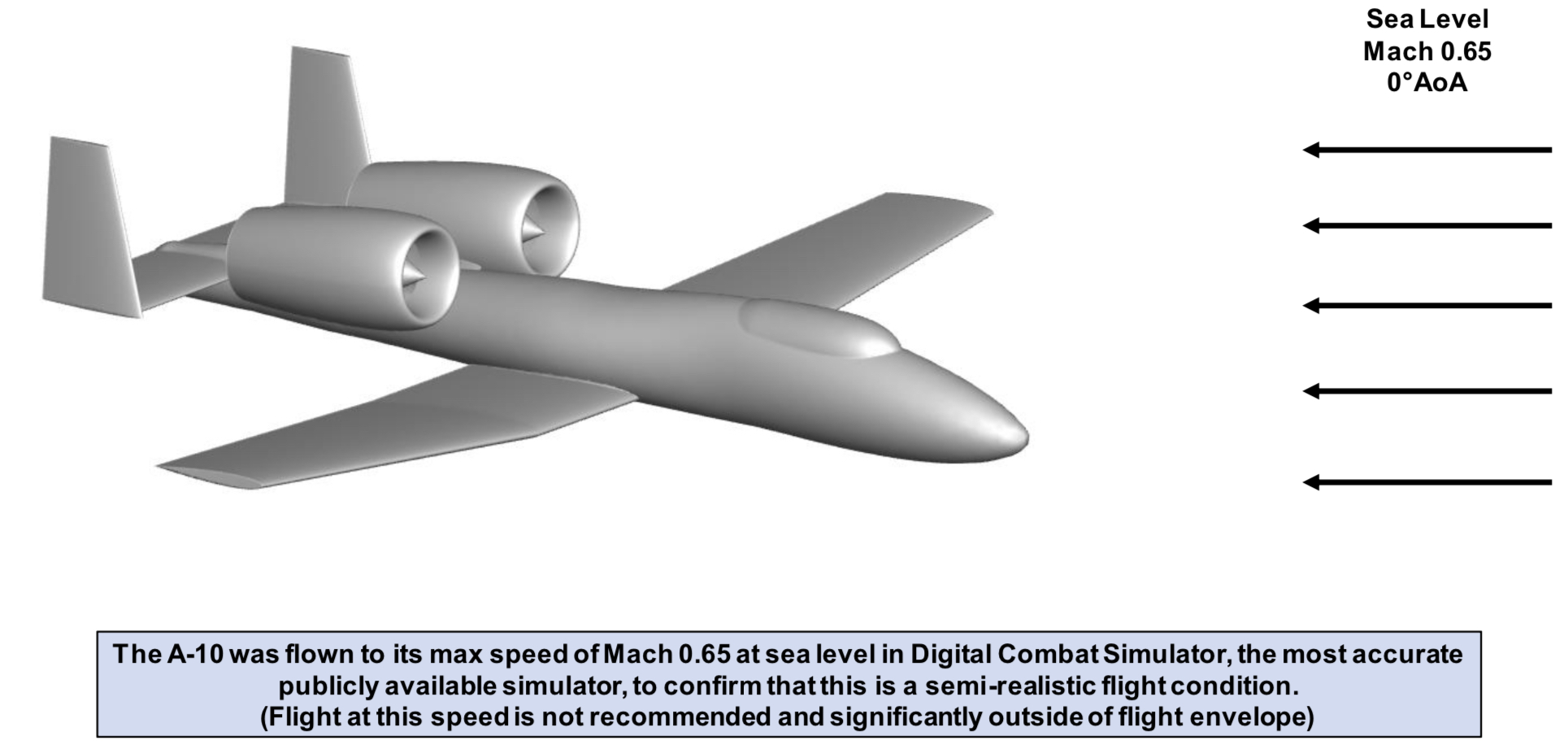}
\caption{the flow conditions used to simulate the flow over the aircraft models in our work.}
\label{fig:cfd}
\end{figure}

After obtaining the CAD surface classifications through voting, we use expert-guided rules to generate meshes automatically for the CFD simulation under specific flow conditions and simulation types~(as shown in Figure \ref{fig:motivation}). These rules would be specific to the CFD simulation type and flow conditions. The primary idea in our work is that a database of rules would be established for each kind of simulation type, and equations will be established for specific mesh settings to be adjusted based on flow conditions. In our work, we focus on a specific setup for CFD simulations and use rules established by experts for mesh generation. Specifically, we develop rules for simulations intended to resolve unsteady flow structure in wake regions for aero-optical work. Furthermore, the mesh generated is meant to be simulated in the FUN3D CFD solver~(second-order time accurate simulation), with Detached Eddy Simulation~(DES) turbulence model. For this work, we limit the scope of work to one specific flow condition as well. This is done to enable us to demonstrate a proof of concept in this paper, and future work will generalize the rules to more situations. The flow conditions we develop our rules around are flight at sea level, at Mach 0.65, and Angle of Attack~(AoA) $0^\circ$~(see figure \ref{fig:cfd}). We also resize airplane models to have a length of 30m. We then develop rules~(based on Pointwise meshing software settings) for different aircraft parts to automate the mesh generation process. We do not go over the specific rules in this section to improve the readability of the paper. These rules are specified in appendix ~\ref{sec:expert-rules}. Although these rules are established automation of the process is not explicitly discussed further in this paper, as mesh settings databases are established by experts at the Lincoln laboratories, and the sensitive nature of their work limits the amount of information that can be discussed publicly. As such, we only provide the details for the above-mentioned flight conditions and see the process of automation as a black box that involves providing surface predictions to experts who can automate the meshing in their desired meshing software closing the loop of expert-guidance integration~(see figure ~\ref{fig:motivation}).

An important thing to note is that the settings that must be applied to the wing are the most refined, followed by the stabilizers, and finally the fuselage has the least refinements. As such when a majority is not reached in CAD surface classification~(see section ~\ref{sec:cadseg}) priority is given to the classification that requires the most refined mesh.

\section{Dataset}
So far, we have discussed a framework for mesh classification and how we can map these predictions to CAD files and use expert guidance to generate CFD meshes. However, there are no publicly available datasets for the purpose of our work. As such to train our model we needed to create a new dataset of airplane models which meet the following criteria. First, the models used for training must be real and accurate aircraft designs and not include examples of toy aircraft and concept designs which are common in datasets such as ShapeNet. This is because, in many of these cases, the features in the models are not indicative of realistic aircraft designs and therefore, may confuse the model and reduce generalizability to real models, and bias the model towards the features observed in unrealistic samples. The other criterion is that the models in the dataset must have segmentation labels pertaining to different aircraft parts, specifically the wing, stabilizers, fuselage, and engine as these are the parts of the aircraft that are important for mesh settings. The third criterion is that the models must include accurate CAD representations associated with them as well since we need to classify the CAD surfaces to apply the mesh settings properly in meshing software.

To create a model that is practically useful for CFD simulations at Lincoln laboratories, we created a dataset of realistic aircraft using Open Vehicle Sketch Pad~(OpenVSP)~\cite{doi:10.2514/6.2022-0004}. OpenVSP, initially developed by NASA, is an open-source tool for creating parametric aircraft geometry that enables the engineering analysis of those models. 
To create the dataset, we use a few selected aircraft designs from OpenVSP and then apply parametric sampling within OpenVSP and data augmentation techniques outside of OpenVSP. In the following sections, we describe how we went about doing this in a more detailed manner.

\subsection{Dataset Backbone}
\begin{figure}[h]
\centering\includegraphics[width=\linewidth]{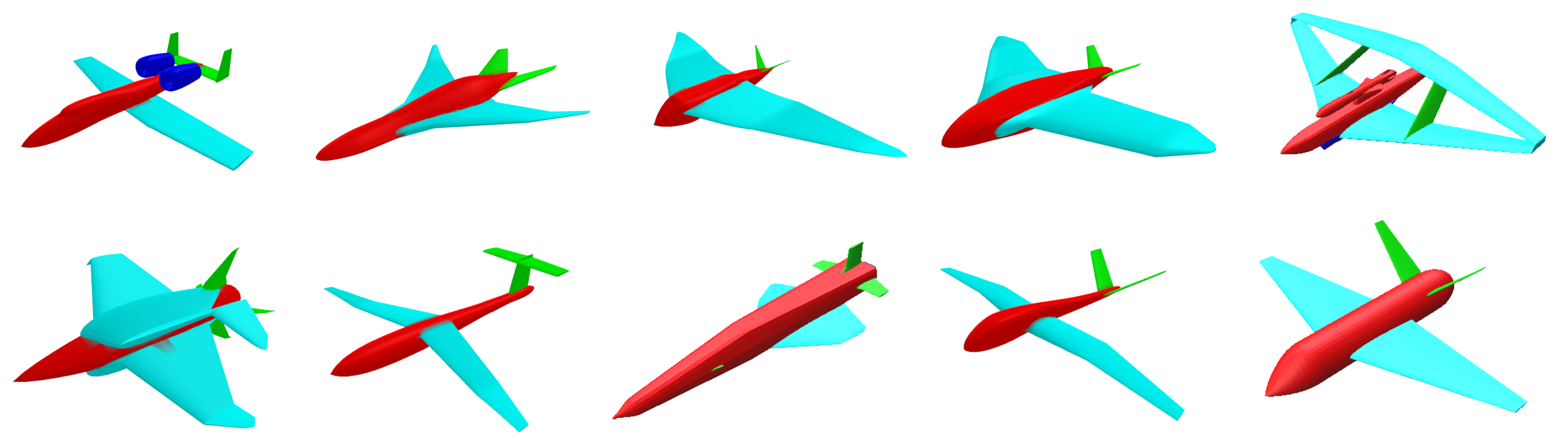}
\caption{The dataset is derived from ten base airplane models, each belonging to a different aircraft type. The models were provided to us by CFD experts, who ensured that they cover a wide range of aerodynamic designs.}
\label{fig:basedata}
\end{figure}

As discussed above, we chose OpenVSP to create a dataset of realistic aircraft models, as their website contains many aerodynamically accurate airplane models within their software, which are made publicly available under the name VSP Hanger\footnote{https://hangar.openvsp.org/}. However, not all of these models can be used easily, as we will see. Before we can use any given model, we need to ensure that we can generate valid geometries for each model and label the different parts of the aircraft. This is a crucial part of the dataset curation, and as a result, we attempt to automate the process as much as possible to enable future dataset expansion.

% The first thing that we need to figure out is an approach for labeling aircraft parts. Withing OpenVSP, there exists a feature that allows users to assign different parts of the geometry to different sets. This conveniently lends itself to our goal of airplane model segmentation. All we need to do is to go through each model and manually assign different parts of the airplane geometry to separate sets. The way we approach this is by assigning each geometric feature to a set and creating lists of sets for each aircraft's fuselage, wing, stabilizers, and engines. Then we extract separate meshes for each set in the OpenVSP model and combine them using MeshLab~\cite{meshlab} boolean operations so that the resulting mesh is water-tight. Since we started with separate meshes for each geometric feature and MeshLab can track the association of each face to the original meshes during boolean operations, we also keep track of which OpenVSP set each face in the mesh belongs to, and using this information and our manual labeling we can generate meshes that are labeled. Given that the dataset is not the primary focus of this work, we will discuss any intricate details of this process here and make the code we use for the processing of the data and the data itself publicly available and refer interested readers to our code and data for further detail.

The initial task is to develop a labeling method for different parts of the aircraft. OpenVSP has a feature that enables users to allocate various parts of the geometry to different sets, which is useful for our goal of segmenting airplane models. We manually assign different parts of the aircraft geometry to various sets, creating lists of sets for each aircraft's fuselage, wing, stabilizers, and engines. We extract separate meshes for each set in the OpenVSP model and use MeshLab's~\cite{meshlab} boolean operations to combine them, resulting in a water-tight mesh. As we keep track of which OpenVSP set each face in the mesh belongs to during boolean operations, we can generate labeled meshes using this information and our manual labeling. We will not delve into the intricacies of this process since the dataset is not the primary focus of our work. However, we will make the code we use for data processing and the data itself publicly available and refer interested readers to our code and data for further detail.

The final note on the data is that given the process of manual labeling and verification of geometric validity is manual and labor intensive at this moment; our dataset is limited to ten accurate aircraft models selected by CFD experts at Lincoln laboratories, which are displayed in figure ~\ref{fig:basedata}. Despite being a small number of models, there is significant variation and diversity in the dataset~(figure ~\ref{fig:basedata}) to capture many different types of aircraft designs. However, this small size presents a major challenge when it comes to applying large deep-learning models, as training a model on eight aircraft models~(assuming two models are set aside for validation) is almost certainly going to lead to overfitting. To overcome this, we propose an augmentation approach which we discuss in the following section. 

\subsection{Working With Small Datasets In Data-Driven Models}
\begin{figure}[h]
\centering\includegraphics[width=\linewidth]{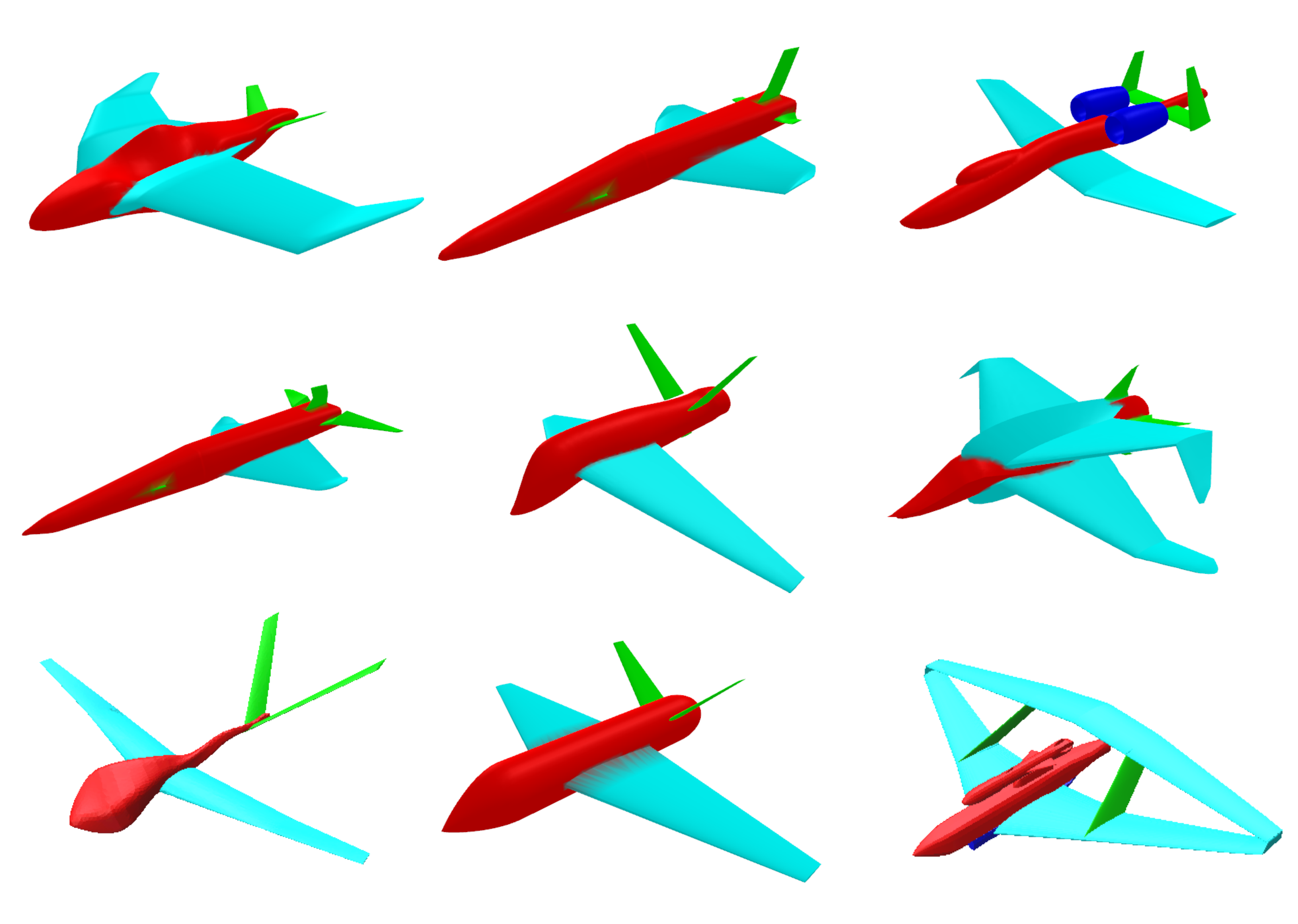}
\caption{Shown above are nine random airplane models from the augmented dataset, which illustrates the significant diversity achieved by changing design parameters. It is important to note that the effects of remeshing are not visible in these images.}
\label{fig:fulldata}
\end{figure}

Since the small size of the dataset does not lend itself to deep learning, we employ some data-level augmentation, which is applied at the data-gathering stage rather than later during training. In our dataset, we employ two strategies for augmentation: design alterations, and 
mesh refinement alteration.

For design alteration, we use OpenVSP's Python API. First, we manually analyze the design parameters~(geometric parameters) in the OpenVSP graphical interface and determine appropriate design parameters that can be altered without breaking the models and the range of acceptable values for said parameters. Once the parameters are determined, we use the OpenVSP API to create randomly altered versions of each of the ten models in our dataset by randomly sampling the values of the design parameters from a uniform distribution of acceptable values. We then generate 20 variations of each of our original models extending the dataset to 210 samples.

One important thing to mention, aside from the limited size of the dataset, is the fact that the mesh generated by OpenVSP has specific patterns in different parts of the aircraft, which could be exploited by a machine learning model to overfit these patterns. Consequently, it is important to ensure the model does not develop a bias towards such patterns. Remeshing the OpenVSP-generated meshes will enable two things for us. The first is that by remeshing at different mesh fidelities, we will augment the dataset and increase the size of the dataset, which is crucial to the success of any learning-based method. The second benefit of this would be that the model cannot be biased toward a specific mesh pattern generated by OpenVSP. To achieve both of these goals, we pick an unbiased uniform remeshing algorithm, isotropic explicit remeshing~\cite{hoppe1993mesh,meshlab}, to remesh each model into five different mesh element sizes and fairly uniform elements, which remove the bias created by software-specific patterns in the mesh. The results of the aforementioned augmentations are visualized in figure ~\ref{fig:fulldata}.

\section{Results and Discussion}
So far, we have established the details of our approach. In this section, we report the results of applying the method to the dataset and compare the performance of our model to other existing state-of-the-art approaches. But before we continue, it is important to establish the training details of our model and augmentation layer.

\subsection{Model Configuration}
To demonstrate the efficacy of our approach in mesh segmentation, we compare our method to two state-of-the-art methods, PointMLP~\cite{ma2022rethinking} and PointNet++~\cite{Qi_2017_ICCV}, that outperform most other models or closely match their performance. These models serve as a benchmark for state-of-the-art. In both of these models, to obtain mesh face classifications we simply add the model's outputs for each vertex on a face and use that as the predictions for the face labels. We also train a naive implementation of a simple graph neural network with six layers of the graph convolution(GCN)~\cite{kipf2016semi} layers as the naive baseline. This model serves as a baseline for the simplest possible approach and is trained to demonstrate the notable gap that would exist without better model architectures. For PointNet++ and PointMLP, we use the code provided by the original authors of the paper and only make the final prediction modification without changing anything in their work or model configuration. For details of the naive baseline, we refer you to our code which will be made publicly available upon the acceptance of this paper.

For our model, we use a $\gamma$ value of 0.1 in our loss~(based on Eqn. ~\ref{eqn:5}), and for the active augmentation layer, we use target parameters of $\xi_1=\pi/6$, $\xi_2=0.001$, $\xi_3=0.2$, $\xi_4=0.4$, and $\xi_5=0.15$. We find that in our experiments, these values yield the best generalizability for our model configuration. We train our model based on the loss function defined in Eqn. ~\ref{eqn:5} and optimize our model using the Adam optimizer with a learning rate that starts at $10^-4$ and decays 15 times during training at equal intervals and at a rate of 0.65. For further details of the size of layers in our architecture and smaller implementation notes, we refer you to our code which will be made publicly available upon the acceptance of this work. To make the comparisons fair, we train all models for 200 epochs and use the original authors' code for the above benchmark models.

\subsection{Mesh Segmentation Results}
\begin{figure}[h]
\centering\includegraphics[width=\linewidth]{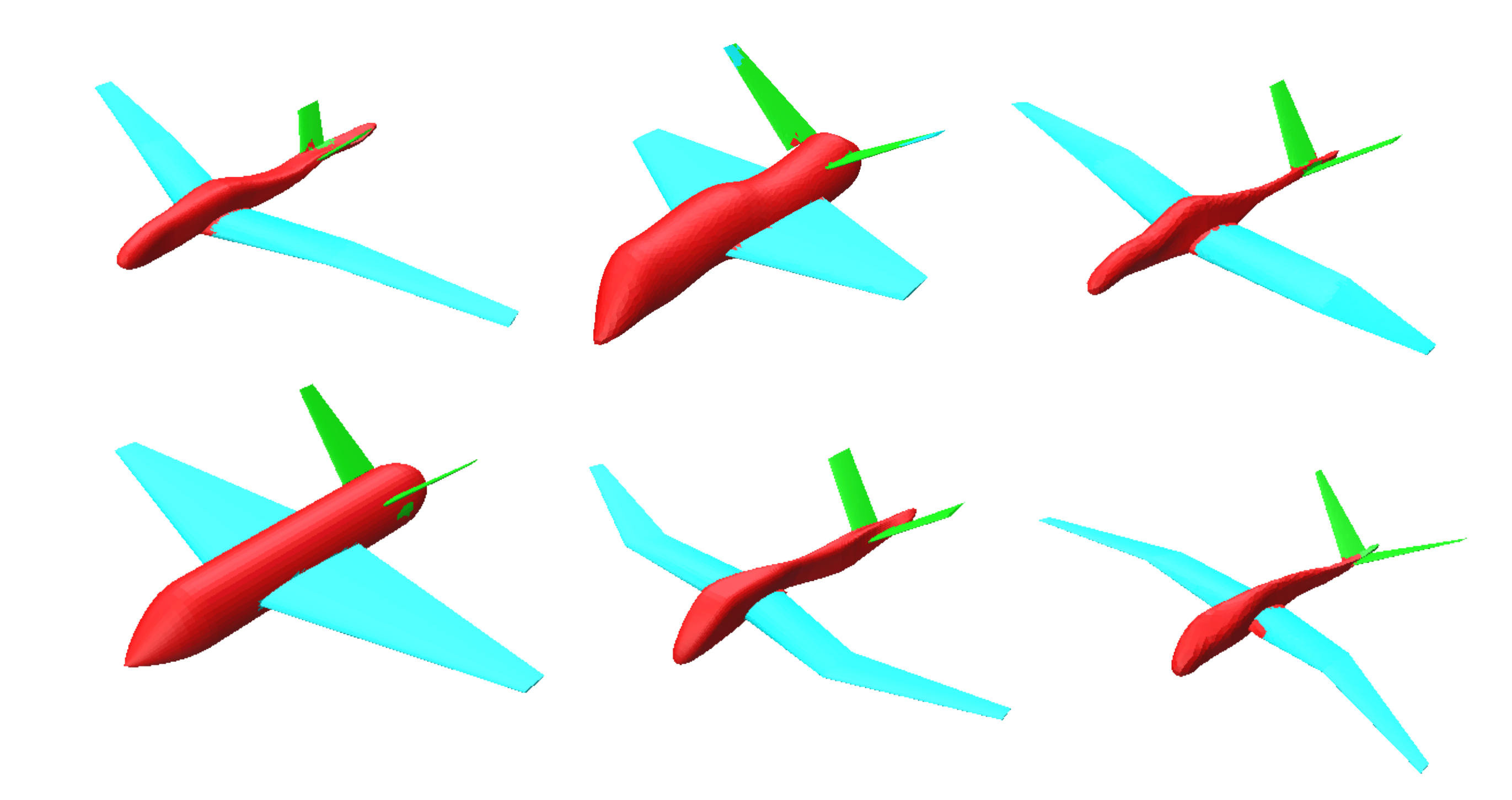}
\caption{We present six sample segmentation predictions generated by our model on previously unseen validation data. The high accuracy of our model on this dataset is apparent from the visual results. Notably, the model correctly classifies the majority of faces associated with each CAD surface, which enables even more accurate CAD surface predictions.}
\label{fig:visres}
\end{figure}

As mentioned above, we train all models on our dataset. For training, we use 840 models ~(based on the first eight in figure ~\ref{fig:basedata}) and leave 210 models generated based on the last two aircraft~(The two models at the bottom right of figure ~\ref{fig:basedata}) for testing and validation. In figure ~\ref{fig:visres}, we visually demonstrate the mesh segmentation predictions made by our model. It is impressive that most of the faces on the meshes are predicted correctly. More importantly, this kind of accuracy will only be amplified when it comes to mapping face predictions to CAD surfaces, as it is much more unlikely that the majority of faces associated with CAD surfaces are incorrectly classified, leading to an even more accurate CAD surface classification.

Besides visual verification of the model's performance, it is important to quantify the accuracy of the model and compare it to state-of-the-art. As such, we train our model and competing models ten times each and record the means value of the best validation accuracy~(percentage of correctly classified faces based on the model's top guess) for each experiment and model and report the results in Table ~\ref{tab:acc}. As evident, our model outperforms all models, including the very recently published and powerful PointMLP, which has established itself as the go-to model for 3D data. Besides that, we do observe that PointNet++ and GCN perform poorly, which is expected as PointMLP is an improvement on the concepts introduced in PointNet++, and the naive GCN model, as expected performs much worse than all other models proving that the baseline of what the simplest possible approach leaves a lot to be desired.

\begin{table}[h]
\caption[Table]{Comparison of the mean accuracy of different models on the validation data. The values after $\pm$ show the standard deviation across ten training runs.}\label{tab:acc}
\centering{%
\begin{tabular}{llr}
\toprule
Model & Top 1 Accuracy (\%)\\
\midrule
PointNet++ & 89.14 $\pm$ 0.25\\
PointMLP & 93.92 $\pm$ 0.15\\
Naive GCN & 81.23 $\pm$ 1.06\\
\midrule
Our Model (w/ Augmentation Layer) & \textbf{96.65 $\pm$ 0.12}\\
Our Model (w/o Augmentation Layer) & 92.35 $\pm$ 0.97\\
\bottomrule
\end{tabular}
}
\end{table}

\subsection{Conformal Predictions Supported CAD Surface Classification}

\begin{table*}[h]
\caption[Table]{Comparison of the accuracy of different models on aircraft surface predictions on the validation data.}\label{tab:surface}
\centering{%
\begin{tabular}{@{}lcccc@{}}
\toprule
Model & \# Incorrect Surfaces & \# Under Refined Surfaces & \# Over Refined Surfaces & Accuracy (\%)\\
\midrule
PointNet++ &23 &  9 & 14 & 97.26\\
PointMLP & 13 & 5& 8 & 98.45\\
Naive GCN & 42 & 9 & 33 & 95.00\\
Naive GCN + Conformal & 116 & \textbf{0} & 116 & 86.19\\
\midrule
Our Model & \textbf{7} & 2 & 5 & \textbf{99.17}\\
Our Model + Conformal & \textbf{7} & \textbf{0} & 7 & \textbf{99.17}\\
\bottomrule
\end{tabular}
}
\end{table*}

For CAD surface classification we take the best-trained model for each method and implement a simple voting algorithm to classify CAD surfaces. This algorithm is very similar to our CAD surface classification~(see figure ~\ref{fig:surface_conformal}) except without conformal predictions such that each face only votes for the top guess of the model. For CAD surface classification we record the number of misclassified aircraft parts (combination of CAD surfaces associated with each part of the CAD file) in the validation data for each approach. Furthermore, we calculate the number of misclassifications that would lead to a more refined mesh than necessary~(ultimately not failing in simulation) and the number of misclassifications that lead to under-refinement, which can be problematic in simulation. We present our results in table ~\ref{tab:surface}. The first observation in these results is the fact that our model which has the most accurate predictions on the mesh as we saw in the prior section, leads to the fewest misclassified surfaces compared to all other models. 

The most notable observation is that using the model's predictions directly without the conformal predictions approach leads to under-refined surfaces, which can affect the simulation efficacy and result in inaccurate solutions, ultimately a failure. This demonstrates the importance of applying conformal predictions to quantify uncertainty and apply the most conservative mesh settings when the model is faced with uncertainty in surface classification. More importantly, conformal predictions enable uncertainty quantification with marginal statistical guarantees, which is a very robust approach, and guarantees under refinement in the mesh will be avoided to a measurable extent which can be set by the user depending on the sensitivity of their work. Another significant observation is that adding the more conservative conformal predictions with $95\%$ marginal guarantee has not caused more over-refinement, which is a testament to the accuracy of our model which is out of the box is good enough that even with the conservative conformal predictions remains accurate. On the contrary, we see that when conformal predictions are applied to models with low accuracies, such as the GCN baseline~(see table ~\ref{tab:surface}), the conformal predictions fail to provide the necessary guarantee easily, leading to much more over-refinement and an overly conservative CFD mesh. Still, it is impressive that conformal prediction still sees no failed CFD mesh with under-refinement even when the model is inaccurate~(See GCN+Conformal in table ~\ref{tab:surface}). To the best of our knowledge, this work is the first demonstration of model-agnostic conformal predictions in the engineering design community, and we hope that our results will inspire further research in different engineering applications to improve deep learning model robustness using a conformal prediction-based approach.

\subsection{Automated Meshing: A Case Study}

\begin{figure}[h!]
\centering\includegraphics[width=\linewidth]{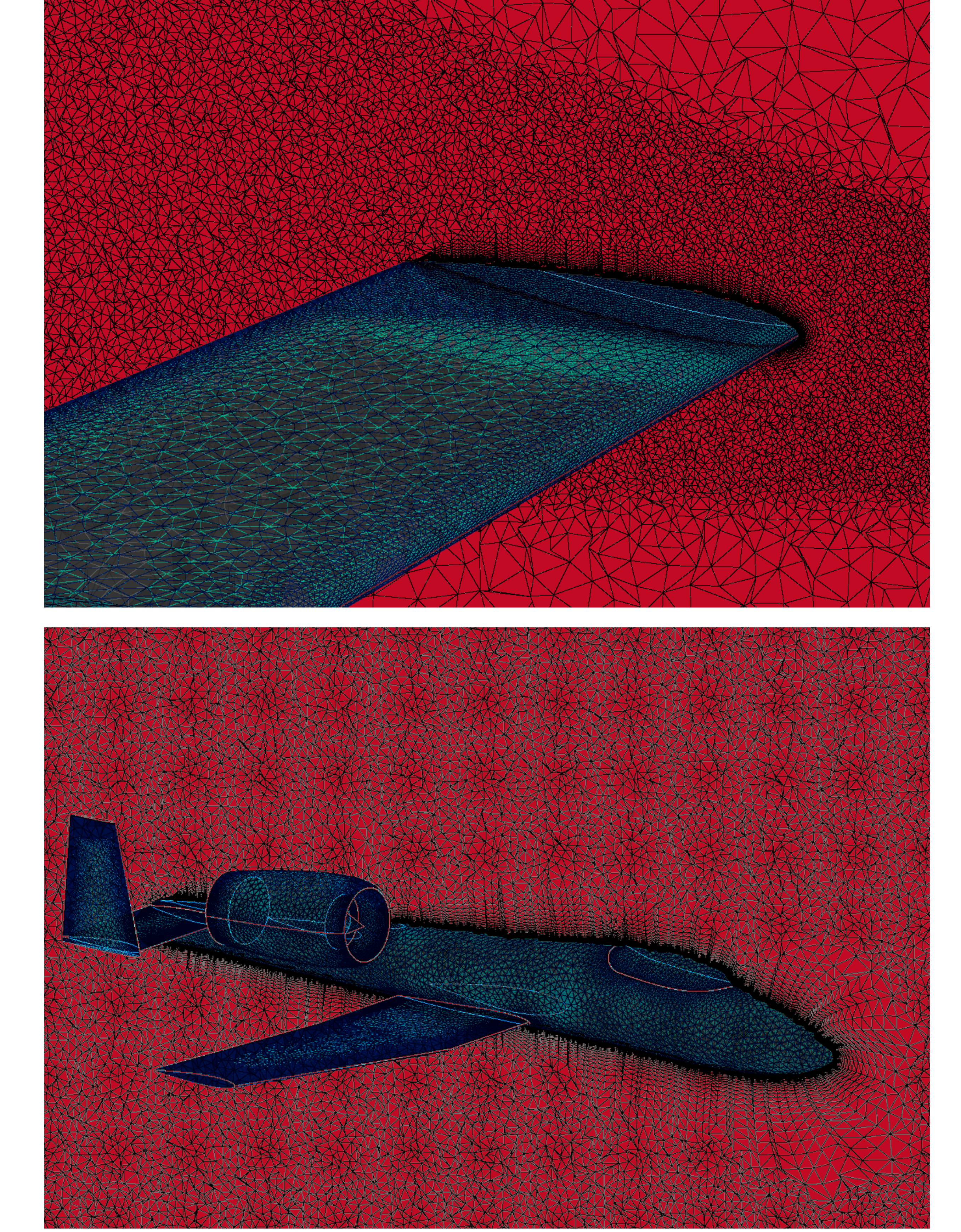}
\caption{These images illustrate how the application of different meshing rules can result in varying levels of refinement in different areas of the model. Specifically, the top two images show how our meshing rules lead to more refinement around the wings, while less refinement is applied around the fuselage where flow patterns are simpler.}
\label{fig:meshcs}
\end{figure}

So far, we have demonstrated that our model is very effective at extracting the necessary geometric insights needed by experts to automate the meshing process. We also discussed the specific settings that can be applied for a specific flight condition. To further showcase the potential of our methods, we conduct a case study on one aircraft model. Specifically, we generate the mesh for the aircraft model resembling the A-10 warthog~(top left aircraft in figure ~\ref{fig:basedata}). We use the expert rules for the flight conditions discussed in section ~\ref{sec:meshingrules}. The results of the rule-based meshing are shown in figure ~\ref{fig:meshcs}. As we discussed, the flow patterns around the wings are the most intricate and require significant refinement compared to the fuselage, and we can see that given the correct classification, the rule-based meshing successfully refined the mesh around the wing while keeping the mesh refinement lower around the fuselage of the aircraft. In this way, the meshing process for this aircraft can be automated with little input or effort from the users without significant added cost on the solution time. These results show that expert-guided automated meshing for aircraft can be of great value for designers and perhaps provide a viable alternative to very expensive adaptive remeshing approaches or even serve as a great initialization to adaptive remeshing leading to significant time saves in both the meshing process itself and the simulation time. 

Finally, to verify that this generated mesh is capable of capturing the necessary details in the simulation, we run the simulation for the generated mesh with the flow conditions mentioned in section \ref{sec:meshingrules}. We observe that the solver converges successfully and the resulting simulations capture the details of the flow intended by the analysis. The efficacy of the meshing step is verified through comparative visualizations of the flow contours along the wing tips, vertical stabilizers, and fuselage where higher detail is required.  Specifically, we see that the simulation results produce the boundary layer flow fields and trailing wake structures expected for this condition, and coefficients of drag and lift adequately converge to our benchmark values of 0.15 and -0.30, respectively.  This occurs after 3 solutions (2 refinement cycles) for the adaptive meshing approach and a single solution for our intelligent meshing approach~(for further details see appendix \ref{sec:adaptive}).  We observe that the total time to a complete simulation using our method is 38 minutes as compared to 210 minutes for adaptive remeshing~(table \ref{tab:time}). This shows that expert-guided refinements are crucial for efficient flow simulations and accurate results.

\begin{figure*}[h]
\centering\includegraphics[width=\linewidth]{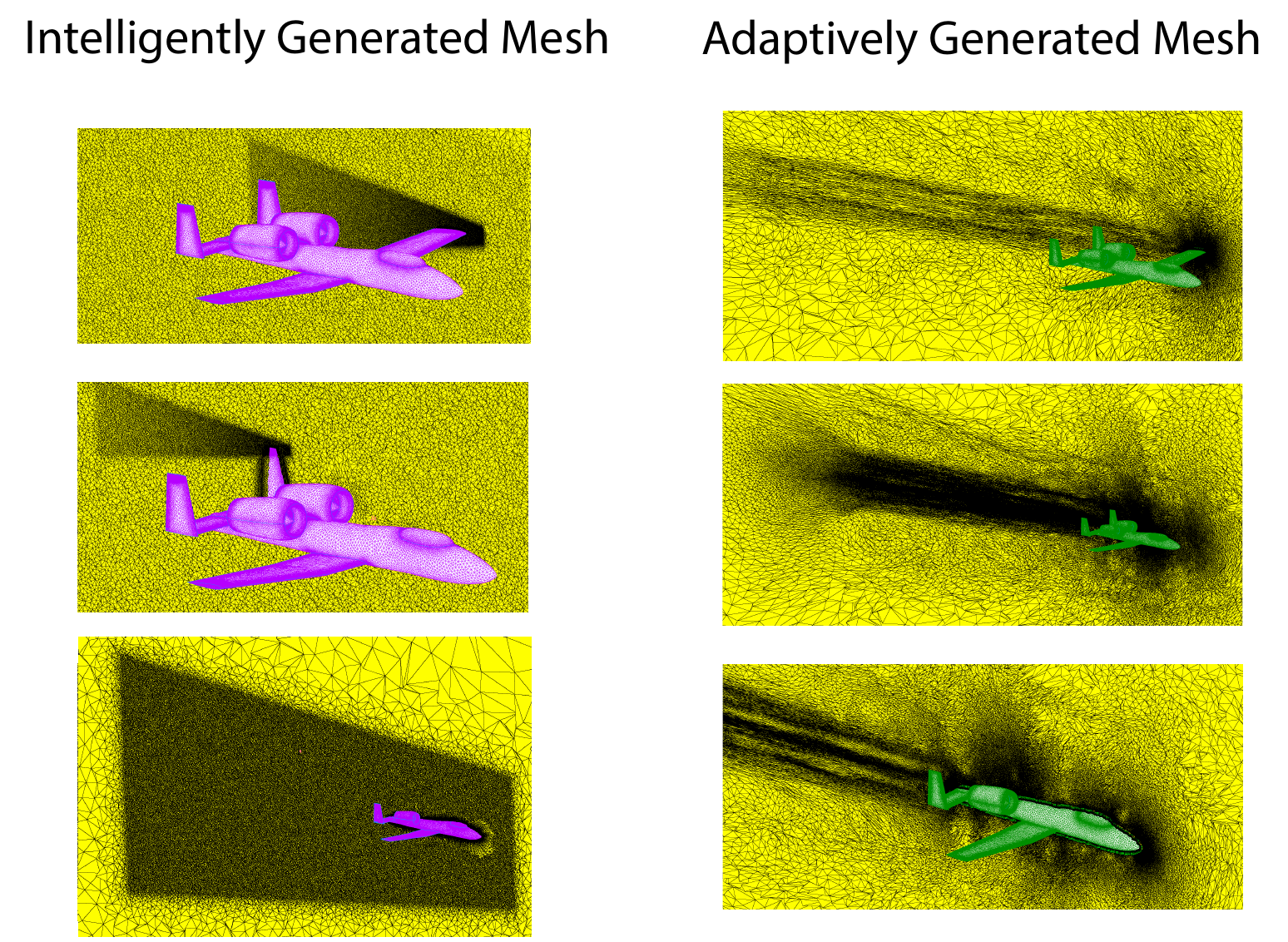}
\caption{We compare our approach for meshing to adaptive meshing in this figure. it can be seen that our intelligently generated mesh has the necessary refinements identified by adaptive remeshing, while not requiring multiple iterations of simulations to be performed.}
\label{fig:adaptive}
\end{figure*}

\begin{table}[h]
\caption[Table]{Comparison of time required to complete a simulation with appropriate flow fields obtained by the simulation algorithm.}\label{tab:time}
\centering{%
\begin{tabular}{llr}
\toprule
Meshing Approach & Total Time\\
\midrule
Adaptive Meshing (3 iterations needed) & 210 minutes\\
\textbf{Intelligent Meshing} & \textbf{38 minutes}\\
\bottomrule
\end{tabular}
}
\end{table}

Furthermore, we observe that expert rules directly address the critical physics of the model and intelligently generate a mesh that can capture the necessary details for further analysis of the simulation while adaptive remeshing must perform multiple, computationally expensive simulations and often still fails to capture sufficient detail in regions of interest.  This is due to the fact that generalized adaptive meshing methods typically focus on large volume convergence of the simulation rather than targeted volumes for down-line analysis by experts. An example of this is the boundary layer prisms in the mesh around the surface of the airplane which will not be refined adaptively (see figure \ref{fig:adaptive}). We compare the mesh generated by the two approaches in figure \ref{fig:adaptive}. As it can be seen the intelligently generated mesh has refinements in places where the adaptive remeshing has identified as regions that require further refinement. However, the intelligently generated mesh seems over-refined when compared to adaptively refined mesh. Despite this given the fact that adaptively refining the mesh involves simulating the problem multiple times (once for each iteration of adaptive remeshing) and the fact that the computational cost of simulation is significantly more than meshing, we see that our intelligently generated mesh is significantly faster overall when the entire process is of simulation is considered.

\section{Design Implications of intelligent meshing}
As we discussed prior CFD simulations play a major role in the design process for many different engineering applications, and the meshing step in performing such simulations  can be rather frustrating and a major limiting factor in many cases. This is because proper meshing for successful and accurate CFD simulation requires significant experience and expert knowledge, which can present a major bottleneck for designers of such systems. This can be mitigated to a great degree by automating the meshing process as the simulation setup is problem dependant and fairly straightforward to set up, however, successful meshing is a more abstract part of the process, and as such our approach presents a major improvement in the overall process, by automating this abstract process. The alternative to our approach would be to perform adaptive meshing, however, this can lead to more than 5 times the time spent on the overall process. This means that our proposed scheme can accelerate the design process for a variety of engineering design applications by a significant amount and allow designers to explore the design space more efficiently. Moreover, this kind of acceleration enables significantly more iterations in the design process which can lead to more optimal design and even enable faster implementations of optimization algorithms that rely on CFD simulations such as the ones studied in this paper.

\section{Limitations and Future Work}
While our methods have been successful, there are areas in which we aim to develop and improve in the future. Currently, our approach is limited to rules based solely on 3D model segmentation, meaning that experts are only able to make rules based on surface types. In the future, we plan to collaborate with CFD experts to create more advanced rules for meshing by identifying other relevant features of 3D models, such as engine inlet and outlet, trailing edges of lift-generating surfaces, and more. This will enable expert guidance to become even more insightful, leading to a more effective and highly optimal mesh generation scheme. We believe that our GNN and point-based model will be able to handle this challenge with ease, and we plan to take this step in the near future.

Another area of improvement for our work would be to expand our dataset to encompass a much larger set of models. This will lead to a more powerful and accurate segmentation model, significantly improving the overall approach. As such, we plan to continue adding models to our dataset in the future and make the data publicly available as a living dataset that will receive updates with more models being added over time.

At this stage, our approach relies on third-party meshing software for CFD mesh generation. However, we believe that a more independent CFD mesh generation pipeline would provide our model with more freedom and enable expert guidance to be significantly more detailed. Therefore, significant effort must be devoted to developing such a pipeline to enable methods like ours to become even more versatile in the future.

\section{Conclusion}

We propose an expert-guided CFD meshing method that combines graph neural networks and expert heuristics to generate CFD meshes that not only improve numerical solver convergence but also capture the necessary details of flow patterns in areas with complex flow patterns. To develop our model, we introduce a novel deep-learning approach that combines the strengths of point-based models and mesh-based GNNs. Additionally, we publicly release a new annotated dataset and show how an active augmentation approach enables us to train a large deep-learning model on a small dataset without sacrificing generalization. We demonstrate that our approach outperforms the latest state-of-the-art techniques for 3D model segmentation, establishing our model as the new state-of-the-art.

Furthermore, we introduce a mechanism for mapping predictions from a mesh to CAD surfaces and utilize conformal predictions in CAD surface classification. Our model is capable of quantifying uncertainty and handling the uncertainty with a $95\%$ marginal statistical guarantee of correctness. Based on meshing rules established for different surface types, we utilize expert guidance to generate CFD meshes. We show that our model provides the most accurate predictions on CAD surfaces~(over 99\% accurate), and our conformal predictions approach leads to correct uncertainty identification and a more conservative approach to meshing with no under-refined surfaces, effectively enabling failure avoidance in the solver.

Finally, we validate the efficacy of our approach by conducting a case-study simulation of one of the aircraft in the dataset. We demonstrate that given accurate surface predictions, the resulting mesh is appropriately refined and leads to successful simulations by the solver, thus establishing the effectiveness of our approach. We make our code and data publicly available for interested readers to examine. The future work includes developing more advanced rules for meshing and expanding the dataset for a more powerful segmentation model. Additionally, efforts will be given to developing an independent CFD mesh generation pipeline for more freedom and detailed expert guidance.

\section*{Acknowledgments}
The authors acknowledge the Lincoln Laboratory for providing computational resources that have contributed to the research results reported in this paper. We also thank MathWorks for supporting Amin Heyrani Nobari's studies while working on this project.

%Bibliography
\bibliographystyle{unsrt}  
\bibliography{references}  

\begin{thebibliography}{10}

\bibitem{padgan}
Wei Chen and Faez Ahmed.
\newblock {PaDGAN: Learning to Generate High-Quality Novel Designs}.
\newblock {\em Journal of Mechanical Design}, 143(3), 11 2020.
\newblock 031703.

\bibitem{mopadgan}
Wei Chen and Faez Ahmed.
\newblock Mo-padgan: Reparameterizing engineering designs for augmented
  multi-objective optimization.
\newblock {\em Applied Soft Computing}, 113:107909, 2021.

\bibitem{heyrani2021pcdgan}
Amin Heyrani~Nobari, Wei Chen, and Faez Ahmed.
\newblock Pcdgan: A continuous conditional diverse generative adversarial
  network for inverse design.
\newblock In {\em Proceedings of the 27th ACM SIGKDD Conference on Knowledge
  Discovery \& Data Mining}, pages 606--616, 2021.

\bibitem{langroudi2020modeling}
A~Tahadjodi Langroudi, F~Zare Afifi, A~Heyrani Nobari, and AF~Najafi.
\newblock Modeling and numerical investigation on multi-objective design
  improvement of a novel cross-flow lift-based turbine for in-pipe hydro energy
  harvesting applications.
\newblock {\em Energy conversion and management}, 203:112233, 2020.

\bibitem{adaptive_remeshing_book}
Tomasz Plewa, Timur Linde, and V.~Gregory Weirs.
\newblock {\em Adaptive mesh refinement: Theory and applications: Proceedings
  of the chicago workshop on adaptive mesh refinement methods, September 3-5,
  2003}.
\newblock Springer, 2005.

\bibitem{fidkowski2011review}
Krzysztof~J Fidkowski and David~L Darmofal.
\newblock Review of output-based error estimation and mesh adaptation in
  computational fluid dynamics.
\newblock {\em AIAA journal}, 49(4):673--694, 2011.

\bibitem{becker_rannacher_2001}
Roland Becker and Rolf Rannacher.
\newblock An optimal control approach to a posteriori error estimation in
  finite element methods.
\newblock {\em Acta Numerica}, 10:1–102, 2001.

\bibitem{BABUSKA19871}
I.~Babuška and A.~Miller.
\newblock A feedback finite element method with a posteriori error estimation:
  Part i. the finite element method and some basic properties of the a
  posteriori error estimator.
\newblock {\em Computer Methods in Applied Mechanics and Engineering},
  61(1):1--40, 1987.

\bibitem{eriksson_estep_hansbo_johnson_1995}
Kenneth Eriksson, Don Estep, Peter Hansbo, and Claes Johnson.
\newblock Introduction to adaptive methods for differential equations.
\newblock {\em Acta Numerica}, 4:105–158, 1995.

\bibitem{VERFURTH199467}
R.~Verfürth.
\newblock A posteriori error estimation and adaptive mesh-refinement
  techniques.
\newblock {\em Journal of Computational and Applied Mathematics}, 50(1):67--83,
  1994.

\bibitem{10.1007/978-3-642-18775-9_13}
Malte Braack and Alexandre Ern.
\newblock Adaptive computation of reactive flows with local mesh refinement and
  model adaptation.
\newblock In Miloslav Feistauer, V{\'i}t Dolej{\v{s}}{\'i}, Petr Knobloch, and
  Karel Najzar, editors, {\em Numerical Mathematics and Advanced Applications},
  pages 159--168, Berlin, Heidelberg, 2004. Springer Berlin Heidelberg.

\bibitem{regenwetter2022deep}
Lyle Regenwetter, Amin~Heyrani Nobari, and Faez Ahmed.
\newblock Deep generative models in engineering design: A review.
\newblock {\em Journal of Mechanical Design}, 144(7):071704, 2022.

\bibitem{huang2021machine}
Keefe Huang, Moritz Kr{\"u}gener, Alistair Brown, Friedrich Menhorn,
  Hans-Joachim Bungartz, and Dirk Hartmann.
\newblock Machine learning-based optimal mesh generation in computational fluid
  dynamics.
\newblock {\em arXiv preprint arXiv:2102.12923}, 2021.

\bibitem{FIDKOWSKI2021109957}
Krzysztof~J. Fidkowski and Guodong Chen.
\newblock Metric-based, goal-oriented mesh adaptation using machine learning.
\newblock {\em Journal of Computational Physics}, 426:109957, 2021.

\bibitem{pfaff2021learning}
Tobias Pfaff, Meire Fortunato, Alvaro Sanchez-Gonzalez, and Peter Battaglia.
\newblock Learning mesh-based simulation with graph networks.
\newblock In {\em International Conference on Learning Representations}, 2021.

\bibitem{doi:10.1080/01621459.2017.1395341}
Mauricio Sadinle, Jing Lei, and Larry Wasserman.
\newblock Least ambiguous set-valued classifiers with bounded error levels.
\newblock {\em Journal of the American Statistical Association},
  114(525):223--234, 2019.

\bibitem{10.5555/3495724.3496026}
Yaniv Romano, Matteo Sesia, and Emmanuel~J. Cand\`{e}s.
\newblock Classification with valid and adaptive coverage.
\newblock In {\em Proceedings of the 34th International Conference on Neural
  Information Processing Systems}, NIPS'20, Red Hook, NY, USA, 2020. Curran
  Associates Inc.

\bibitem{angelopoulos2020uncertainty}
Anastasios Angelopoulos, Stephen Bates, Jitendra Malik, and Michael~I Jordan.
\newblock Uncertainty sets for image classifiers using conformal prediction.
\newblock {\em arXiv preprint arXiv:2009.14193}, 2020.

\bibitem{Su2015MultiviewCN}
Hang Su, Subhransu Maji, Evangelos Kalogerakis, and Erik~G. Learned-Miller.
\newblock Multi-view convolutional neural networks for 3d shape recognition.
\newblock {\em 2015 IEEE International Conference on Computer Vision (ICCV)},
  pages 945--953, 2015.

\bibitem{kalogerakis20173d}
Evangelos Kalogerakis, Melinos Averkiou, Subhransu Maji, and Siddhartha
  Chaudhuri.
\newblock 3d shape segmentation with projective convolutional networks.
\newblock In {\em proceedings of the IEEE conference on computer vision and
  pattern recognition}, pages 3779--3788, 2017.

\bibitem{7298801}
Zhirong Wu, Shuran Song, Aditya Khosla, Fisher Yu, Linguang Zhang, Xiaoou Tang,
  and Jianxiong Xiao.
\newblock 3d shapenets: A deep representation for volumetric shapes.
\newblock In {\em 2015 IEEE Conference on Computer Vision and Pattern
  Recognition (CVPR)}, pages 1912--1920, 2015.

\bibitem{Brock2016GenerativeAD}
Andrew Brock, Theodore Lim, James~M. Ritchie, and Nick Weston.
\newblock Generative and discriminative voxel modeling with convolutional
  neural networks.
\newblock {\em ArXiv}, abs/1608.04236, 2016.

\bibitem{Tchapmi2017SEGCloudSS}
Lyne~P. Tchapmi, Christopher~Bongsoo Choy, Iro Armeni, JunYoung Gwak, and
  Silvio Savarese.
\newblock Segcloud: Semantic segmentation of 3d point clouds.
\newblock {\em 2017 International Conference on 3D Vision (3DV)}, pages
  537--547, 2017.

\bibitem{Hanocka2018ALIGNetPA}
Rana Hanocka, Noa Fish, Zhenhua Wang, Raja Giryes, Shachar Fleishman, and
  Daniel Cohen-Or.
\newblock Alignet: Partial-shape agnostic alignment via unsupervised learning.
\newblock {\em ACM Trans. Graph.}, 38:1:1--1:14, 2018.

\bibitem{7353481}
Daniel Maturana and Sebastian Scherer.
\newblock Voxnet: A 3d convolutional neural network for real-time object
  recognition.
\newblock In {\em 2015 IEEE/RSJ International Conference on Intelligent Robots
  and Systems (IROS)}, pages 922--928, 2015.

\bibitem{10.1109/TIP.2020.3039378}
Kai Sun, Jiangshe Zhang, Junmin Liu, Ruixuan Yu, and Zengjie Song.
\newblock Drcnn: Dynamic routing convolutional neural network for multi-view 3d
  object recognition.
\newblock {\em Trans. Img. Proc.}, 30:868–877, jan 2021.

\bibitem{Qi_2017_CVPR}
Charles~R. Qi, Hao Su, Kaichun Mo, and Leonidas~J. Guibas.
\newblock Pointnet: Deep learning on point sets for 3d classification and
  segmentation.
\newblock In {\em Proceedings of the IEEE Conference on Computer Vision and
  Pattern Recognition (CVPR)}, July 2017.

\bibitem{NIPS2017_d8bf84be}
Charles~Ruizhongtai Qi, Li~Yi, Hao Su, and Leonidas~J Guibas.
\newblock Pointnet++: Deep hierarchical feature learning on point sets in a
  metric space.
\newblock In I.~Guyon, U.~Von Luxburg, S.~Bengio, H.~Wallach, R.~Fergus,
  S.~Vishwanathan, and R.~Garnett, editors, {\em Advances in Neural Information
  Processing Systems}, volume~30. Curran Associates, Inc., 2017.

\bibitem{dgcnn}
Yue Wang, Yongbin Sun, Ziwei Liu, Sanjay~E. Sarma, Michael~M. Bronstein, and
  Justin~M. Solomon.
\newblock Dynamic graph cnn for learning on point clouds.
\newblock {\em ACM Transactions on Graphics (TOG)}, 2019.

\bibitem{huang2018recurrent}
Qiangui Huang, Weiyue Wang, and Ulrich Neumann.
\newblock Recurrent slice networks for 3d segmentation of point clouds.
\newblock In {\em Proceedings of the IEEE conference on computer vision and
  pattern recognition}, pages 2626--2635, 2018.

\bibitem{NEURIPS2018_f5f8590c}
Yangyan Li, Rui Bu, Mingchao Sun, Wei Wu, Xinhan Di, and Baoquan Chen.
\newblock Pointcnn: Convolution on x-transformed points.
\newblock In S.~Bengio, H.~Wallach, H.~Larochelle, K.~Grauman, N.~Cesa-Bianchi,
  and R.~Garnett, editors, {\em Advances in Neural Information Processing
  Systems}, volume~31. Curran Associates, Inc., 2018.

\bibitem{Wu_2019_CVPR}
Wenxuan Wu, Zhongang Qi, and Li~Fuxin.
\newblock Pointconv: Deep convolutional networks on 3d point clouds.
\newblock In {\em Proceedings of the IEEE/CVF Conference on Computer Vision and
  Pattern Recognition (CVPR)}, June 2019.

\bibitem{ma2022rethinking}
Xu~Ma, Can Qin, Haoxuan You, Haoxi Ran, and Yun Fu.
\newblock Rethinking network design and local geometry in point cloud: A simple
  residual {MLP} framework.
\newblock In {\em International Conference on Learning Representations}, 2022.

\bibitem{Hanocka2019MeshCNNAN}
Rana Hanocka, Amir Hertz, Noa Fish, Raja Giryes, Shachar Fleishman, and Daniel
  Cohen-Or.
\newblock Meshcnn: a network with an edge.
\newblock {\em ACM Transactions on Graphics (TOG)}, 38:1 -- 12, 2019.

\bibitem{8984309}
Chunfeng Lian, Li~Wang, Tai-Hsien Wu, Fan Wang, Pew-Thian Yap, Ching-Chang Ko,
  and Dinggang Shen.
\newblock Deep multi-scale mesh feature learning for automated labeling of raw
  dental surfaces from 3d intraoral scanners.
\newblock {\em IEEE Transactions on Medical Imaging}, 39(7):2440--2450, 2020.

\bibitem{9720214}
Youyi Zheng, Beijia Chen, Yuefan Shen, and Kaidi Shen.
\newblock Teethgnn: Semantic 3d teeth segmentation with graph neural networks.
\newblock {\em IEEE Transactions on Visualization and Computer Graphics}, pages
  1--1, 2022.

\bibitem{WEN2021106250}
Tingxi Wen, Jiafu Zhuang, Yu~Du, Linjie Yang, and Jianfei Xu.
\newblock Dual-sampling attention pooling for graph neural networks on 3d mesh.
\newblock {\em Computer Methods and Programs in Biomedicine}, 208:106250, 2021.

\bibitem{Qi_2017_ICCV}
Xiaojuan Qi, Renjie Liao, Jiaya Jia, Sanja Fidler, and Raquel Urtasun.
\newblock 3d graph neural networks for rgbd semantic segmentation.
\newblock In {\em Proceedings of the IEEE International Conference on Computer
  Vision (ICCV)}, Oct 2017.

\bibitem{stylegan3}
Tero Karras, Miika Aittala, Janne Hellsten, Samuli Laine, Jaakko Lehtinen, and
  Timo Aila.
\newblock Training generative adversarial networks with limited data, 2020.

\bibitem{10.1115/1.4052442}
Amin~Heyrani Nobari, Wei Chen, and Faez Ahmed.
\newblock {Range-Constrained Generative Adversarial Network: Design Synthesis
  Under Constraints Using Conditional Generative Adversarial Networks}.
\newblock {\em Journal of Mechanical Design}, 144(2), 10 2021.
\newblock 021708.

\bibitem{velivckovic2017graph}
Petar Veli{\v{c}}kovi{\'c}, Guillem Cucurull, Arantxa Casanova, Adriana Romero,
  Pietro Lio, and Yoshua Bengio.
\newblock Graph attention networks.
\newblock {\em arXiv preprint arXiv:1710.10903}, 2017.

\bibitem{doi:10.2514/6.2022-0004}
Robert~A. McDonald and James~R. Gloudemans.
\newblock {\em Open Vehicle Sketch Pad: An Open Source Parametric Geometry and
  Analysis Tool for Conceptual Aircraft Design}.

\bibitem{meshlab}
Paolo Cignoni, Marco Callieri, Massimiliano Corsini, Matteo Dellepiane, Fabio
  Ganovelli, and Guido Ranzuglia.
\newblock {MeshLab: an Open-Source Mesh Processing Tool}.
\newblock In Vittorio Scarano, Rosario~De Chiara, and Ugo Erra, editors, {\em
  Eurographics Italian Chapter Conference}. The Eurographics Association, 2008.

\bibitem{hoppe1993mesh}
Hugues Hoppe, Tony DeRose, Tom Duchamp, John McDonald, and Werner Stuetzle.
\newblock Mesh optimization.
\newblock In {\em Proceedings of the 20th annual conference on Computer
  graphics and interactive techniques}, pages 19--26, 1993.

\bibitem{kipf2016semi}
Thomas~N Kipf and Max Welling.
\newblock Semi-supervised classification with graph convolutional networks.
\newblock {\em arXiv preprint arXiv:1609.02907}, 2016.

\end{thebibliography}

\appendix

\section{Expert-Guided Rules For Different Aircraft Surfaces}
\label{sec:expert-rules}
In this section, we present the specific rules provided by experts we use for the flight conditions and software mentioned in section ~\ref{sec:meshingrules}. Below are the specific mesh settings that are applied in PointWise meshing software for different aircraft parts:

\begin{enumerate}
    \item \textbf{Wings:}
    \begin{itemize}
        \item Set surface mesh dimension = 0.05 
        \item Set surface mesh to:  quadrilateral cells dominant 
        \item Set volume mesh to:  hexahedral cells dominant 
        \item Set initial wall spacing = 4.7e-6 
        \item Set Growth rate = 1.1 
        \item Set Collision Buffer = 2.0 
        \item Grow volume cells from surface cells. Check Y+ value...  If > 1, then decrease initial wall spacing proportionally 
    \end{itemize}
    \item \textbf{Stabalizers:}
    \begin{itemize}
        \item Set surface mesh dimension = 0.2  
        \item Set surface mesh to:  quadrilateral cells dominant 
        \item Set volume mesh to:  hexahedral cells dominant 
        \item Set initial wall spacing = 4.7e-6 
        \item Set Growth rate = 1.1 
        \item Set Collision Buffer = 2.0 
        \item Grow volume cells from surface cells. Check Y+ value...  If > 1, then decrease initial wall spacing proportionally 
    \end{itemize}
    \item \textbf{Fuselage:}
    \begin{itemize}
        \item Set surface mesh dimension = 1.0
        \item Set surface mesh to:  quadrilateral cells dominant 
        \item Set volume mesh to:  hexahedral cells dominant 
        \item Set initial wall spacing = 4.7e-6 
        \item Set Growth rate = 1.1 
        \item Set Collision Buffer = 2.0 
        \item Grow volume cells from surface cells. Check Y+ value...  If > 1, then decrease initial wall spacing proportionally 
    \end{itemize}
\end{enumerate}

\section{Further Details On Adaptive Remeshing}
\label{sec:adaptive}
\begin{figure*}[h]
\centering\includegraphics[width=\linewidth]{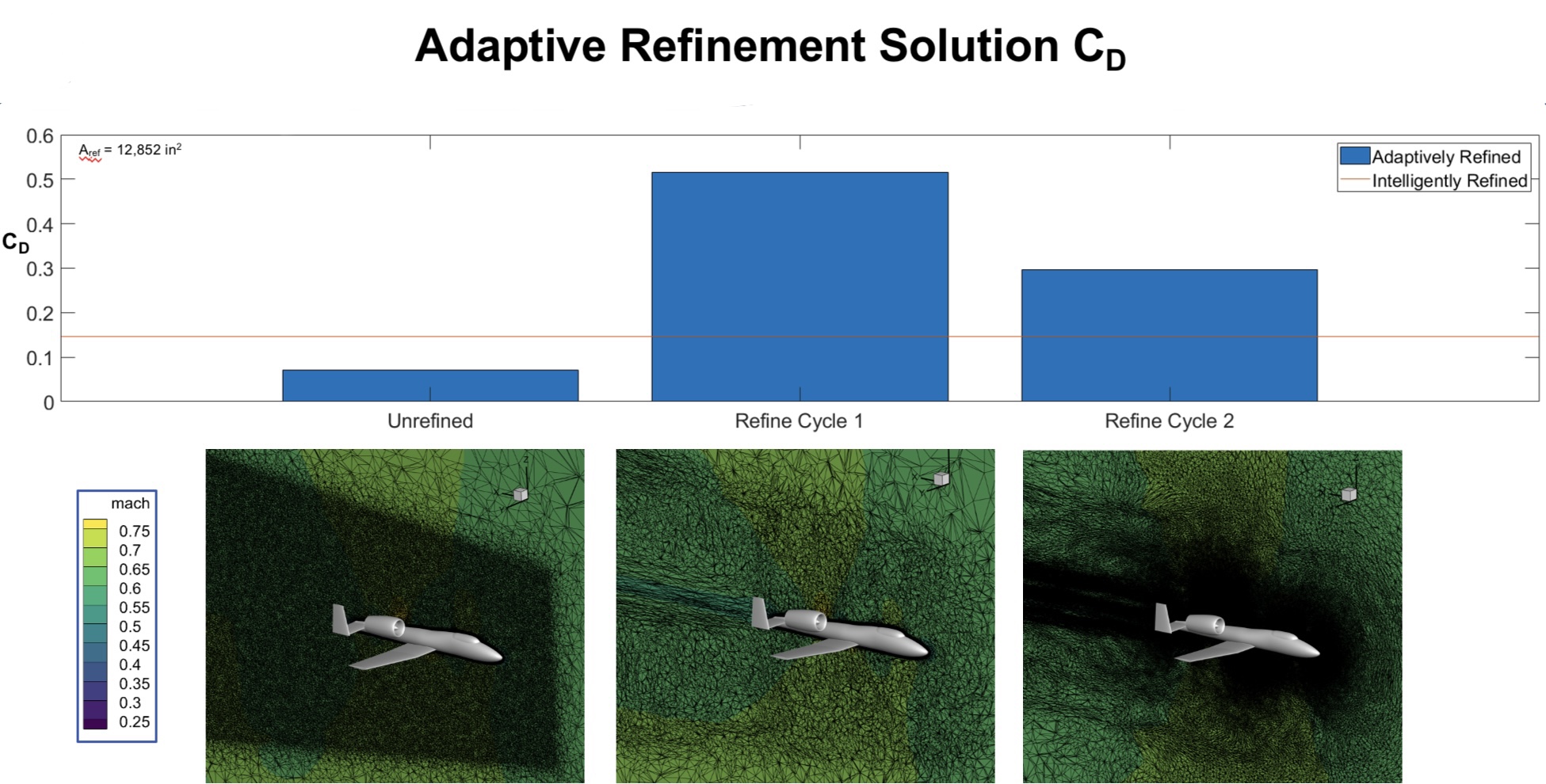}
\includegraphics[width=\linewidth]{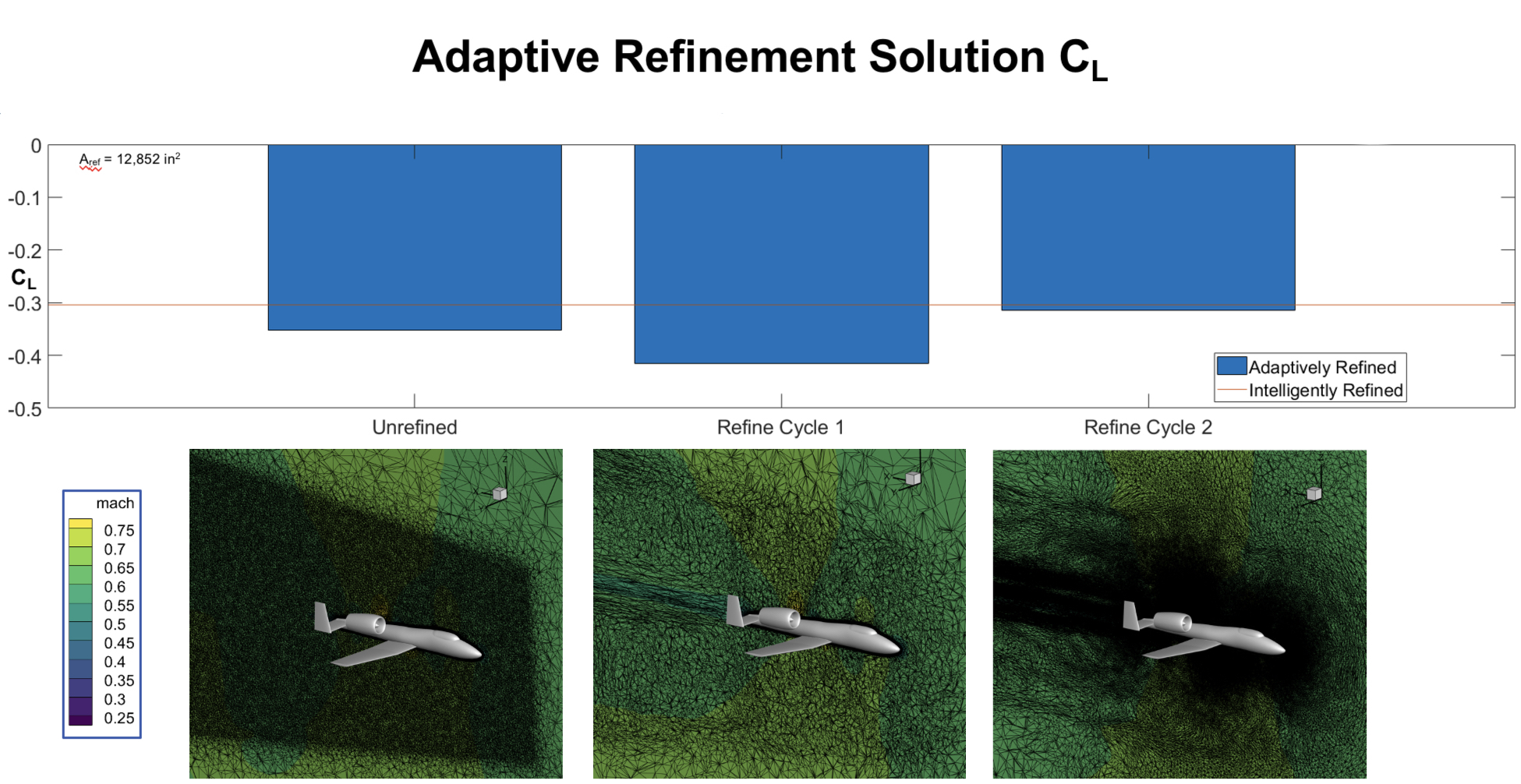}
\caption{Lift and drag coefficient simulation results for the two adaptive remeshing simulations.}
\label{fig:iters}
\end{figure*}

We perform adaptive remeshing until the results of the simulations for coefficients of drag and lift adequately converge to our benchmark values of 0.15 and -0.30, respectively~(see figure \ref{fig:iters}). This is only possible when these benchmark values are available, however, this is a forgiving measure of the performance of adaptive meshing as convergence has to be verified with more iterations in the general case. Regardless, this occurs after 3 solutions (2 refinement cycles) for the adaptive meshing approach and a single solution for our intelligent meshing approach. The results of the simulations for adaptive remeshing are presented here in figure \ref{fig:iters}.

\end{document}